\documentclass[aps,prd,preprint,superscriptaddress,showpacs,floatfix,nobibnotes,nofootinbib]{revtex4}
\usepackage{mathrsfs}

\usepackage{bm}
\usepackage{graphicx}
\usepackage{longtable}
\usepackage{amsmath}
\usepackage{amssymb}

\begin{document}

\title{Mass and width of $T_{c\bar c}(4020)$ in the developed Bethe-Salpeter theory}

\author{Xiaozhao Chen}\email{chen_xzhao@sina.com}
\email[corresponding author]{} \affiliation{Department of Fundamental Courses, Shandong University of Science and Technology, Taian 271019, China}

\author{Xiaofu L\"{u}}
\affiliation{Department of Physics, Sichuan University, Chengdu 610064, China}
\affiliation{Institute of Theoretical Physics, The Chinese Academy of Sciences, Beijing 100080, China}
\affiliation{CCAST (World Laboratory), P.O. Box 8730, Beijing  100080, China}

\author{Xiurong Guo}
 \affiliation{Department of Fundamental Courses, Shandong University of Science and Technology, Taian 271019, China}

\author{Zonghua Shi}
 \affiliation{Department of Fundamental Courses, Shandong University of Science and Technology, Taian 271019, China}

\date{\today}

\begin{abstract}
In experiments exotic meson resonance $T_{c\bar c}(4020)$ lies above the $D^{*}\bar{D}^{*}$ threshold, and in principle one can not explain $T_{c\bar c}(4020)$ as a meson-meson bound state because meson-meson bound state must lie below the $D^{*}\bar{D}^{*}$ threshold. In this work, exotic resonance $T_{c\bar c}(4020)$ is considered as an unstable meson-meson molecular state $D^{*}\bar{D}^{*}$, and the developed Bethe-Salpeter theory for dealing with unstable state is applied to investigate resonance $T_{c\bar c}(4020)$. We calculate the mass and width of unstable meson-meson molecular state $D^{*}\bar{D}^{*}$ in the framework of relativistic quantum field theory and find that this unstable meson-meson molecular state lies above the $D^{*}\bar{D}^{*}$ threshold, which is consistent with experimental values of resonance $T_{c\bar c}(4020)$.
\end{abstract}

\pacs{12.40.Yx, 14.40.Rt, 12.39.Ki}


\maketitle

\newpage

\parindent=20pt

\section{introduction}
The charged charmonium-like state $T_{c\bar c}(4020)$, also named $Z_c(4020)$, has been observed in a series of experiments \cite{X40201,X40202,X40203,X40204,X40205} and its properties are incompatible with a $c\bar c$ structure. Hadronic molecule structure has been proposed to interpret the internal structure of exotic meson resonance $T_{c\bar c}(4020)$ \cite{ms:Guo,ms:Dong,ms:Liu}. In these previous works, exotic resonance $T_{c\bar c}(4020)$ was considered as meson-meson bound state $D^{*}\bar{D}^{*}$ with spin-parity quantum numbers $1^+$ and the standard homogeneous Bethe-Salpeter (BS) equation \cite{BSE:SALPETER1} was used to investigate molecular state. Solving the standard homogeneous BS equation for meson-meson bound state $D^{*}\bar{D}^{*}$, the authors of Refs. \cite{ms:Guo,ms:Dong,ms:Liu} obtained mass and BS wave function. The mass of meson-meson bound state $D^{*}\bar{D}^{*}$ was regarded as mass of exotic meson resonance $T_{c\bar c}(4020)$. However, in experiments exotic resonance $T_{c\bar c}(4020)$ lies above the $D^{*}\bar{D}^{*}$ threshold, and in principle one can not explain $T_{c\bar c}(4020)$ as meson-meson bound state because meson-meson bound state must lie below the $D^{*}\bar{D}^{*}$ threshold. In this paper, we consider exotic meson resonance $T_{c\bar c}(4020)$ as an unstable molecular state $D^{*}\bar{D}^{*}$ with spin-parity quantum numbers $1^+$, and the developed Bethe-Salpeter theory (DBST) \cite{mypaper8} is applied to calculate mass and width of this unstable molecular state. We find that this unstable meson-meson molecular state lies above the $D^{*}\bar{D}^{*}$ threshold.

It is unambiguous that the standard homogeneous Bethe-Salpeter equation (BSE) only describes two-body bound state. Clearly in the standard homogeneous BSE the bound-state mass is a real number and bound state lies below the threshold \cite{BSE:SALPETER1}. In order to describe resonance, one has to develop the standard homogeneous BSE. So far, there are two reasonable and feasible strategies to develop the standard homogeneous BSE for dealing with resonance in the framework of relativistic quantum field theory. One of these two strategies is that the total momentum of two-body system is complex valued, and then the resonance information is contained in homogeneous BSE, as demonstrated in Refs. \cite{BScomplex1,BScomplex2,fourbody2,BScomplex3}. In these works, great efforts have been made to extract the resonance information from homogeneous BSE and contour deformations were employed. If the total momentum is complex valued, one has to devise a suitable integration path to avoid the cuts stemming from the BS kernel. After analytic continuation across the branch cut starting at the threshold, one can obtain the pole describing resonance on the second Riemann sheet. These previous works \cite{BScomplex1,BScomplex2,fourbody2,BScomplex3} provided a feasible method to extract the resonance information from homogeneous BSE. Another strategy to study resonance is DBST \cite{mypaper8}. In this approach, one studies the time evolution of unstable two-body system determined by the total Hamiltonian and then obtains the pole describing resonance on the second Riemann sheet. The numerical calculation in DBST does not involve contour deformations and analytic continuation. Obviously, each strategy to study resonance has its own advantages, and these two strategies are mutually complementary and can serve to validate each other's findings.

In DBST, we suppose that at some given time the unstable two-body system has been prepared to decay and then study the time evolution of this system as determined by the total Hamiltonian. This prepared state can be described by the ground-state BS wave function for vector-vector bound state at the times $t_1=0$ and $t_2=0$. It is necessary to emphasize that solution of homogeneous Bethe-Salpeter equation for two-body bound state only describes the prepared state and it can not describe unstable system. From the effective theory at low energy QCD, we construct the interaction kernel between two quarks in two heavy vector mesons derived from one light-meson exchange and one heavy-meson exchange in the framework of relativistic quantum field theory. Solving BS equation with this interaction kernel, we can obtain the mass and BS wave function for bound state composed of two vector mesons $D^{*+}\bar{D}^{*0}$ with spin-parity quantum numbers $1^+$. After providing the description for the prepared state, we can study the time evolution of the prepared state and obtain the pole corresponding to resonance through the scattering matrix element.

All decay channels of resonance should contribute to its physical mass, and DBST has given a technique for dealing with the dynamics of coupled channels in the framework of relativistic quantum field theory \cite{mypaper8}. For exotic resonance $T_{c\bar c}(4020)^+$, we consider two decay channels $h_c(1P)\pi^+$ and $D^{*+}\bar D^{*0}$. Applying DBST, we calculate the scattering matrix elements between bound states with respect to arbitrary value of the final state energy. It is necessary to note that the total energy of the final state extends over the real interval while the initial state energy is specified \cite{mypaper8}. Then we obtain the correction for energy level of molecular state due to decay channels and find that the unstable meson-meson molecular state $D^{*+}\bar D^{*0}$ lies above the threshold. Finally, the physical mass is used to calculate decay widths of physical resonance $T_{c\bar c}(4020)^+$.

The investigation of the structure of $T_{c\bar c}(4020)$ is of great significance, while the structure has not been determined. Up to now, many theoretical interpretations for this exotic resonance have been proposed. In Refs. \cite{ms:Guo,ms:Dong,ms:Liu}, $T_{c\bar c}(4020)$ was considered as meson-meson bound state $D^{*}\bar{D}^{*}$ and the heavy mesons in molecular state were considered as pointlike objects. To describe exotic meson resonance composed of four quarks, the authors of Refs. \cite{fourbody1,fourbody3,fourbody4} extended the homogeneous Bethe-Salpeter equation for two-body system to deal with four-body system. In these works \cite{fourbody1,fourbody3,fourbody4}, exotic meson resonance was treated as the coupled mixture of meson-meson and diquark-antidiquark states, and the four-body BS equation was used to investigate this coupled mixture. In this paper, we use DBST to investigate unstable meson-meson molecular state. Exotic meson resonance $T_{c\bar c}(4020)$ is considered as an unstable molecular state $D^{*}\bar{D}^{*}$ and heavy mesons $D^{*}$ and $\bar{D}^{*}$ are considered as quark-antiquark bound states, which means that the unstable meson-meson molecular state is considered as an unstable four-body system. After obtaining BS wave function of the prepared state, we study the time evolution of the prepared state composed of four quarks and obtain the pole corresponding to resonance. Obviously, our approach can be extended to the coupled mixture of unstable meson-meson and diquark-antidiquark states, and analogous mix has been investigated in our previous work \cite{mypaper9}. In our future article, the coupled mixture of unstable meson-meson and diquark-antidiquark states will be studied.

The structure of this article is as follows. Section \ref{sec:DBST} gives the developed Bethe-Salpeter theory. In Sec. \ref{sec:BS} we give BS equation for bound state composed of two vector mesons. The mass and BS wave function for bound state composed of two vector mesons $D^{*+}\bar D^{*0}$ with spin-parity quantum numbers $1^+$ are obtained. In Sec. \ref{sec:T-ME} we introduce the matrix elements between bound states with respect to arbitrary value of the final state energy. Two decay channels $h_c(1P)\pi^+$ and $D^{*+}\bar D^{*0}$ are considered. In Sec. \ref{sec:mw} we obtain the physical mass and width for unstable molecular state. Our numerical results are presented in Sec. \ref{sec:nr} and we make some concluding remarks in Sec. \ref{sec:concl}.

\section{the developed Bethe-Salpeter theory}\label{sec:DBST}
To deal with resonance in the framework of relativistic quantum field theory, we considered the time evolution of molecular state as determined by the total Hamiltonian and provided the developed Bethe-Salpeter theory in Refs. \cite{mypaper8,mypaper9}. Because the time evolution of molecular state is determined by the total Hamiltonian, exotic meson resonance should be considered as an unstable meson-meson molecular state. According to the developed Bethe-Salpeter theory for dealing with resonance \cite{mypaper8}, this unstable state has been prepared to decay at given time, and the prepared state can be regarded as a bound state with ground-state energy. Solving BS equation for arbitrary meson-meson bound state, one can obtain the mass $M_0$ and BS wave function $\chi_P(x'_1,x'_2)$ for this bound state with momentum $P=(\mathbf{P},i\sqrt{\mathbf{P}^2+M_0^2})$, where $x_1'=(\mathbf{x}_1',it_1)$ and $x_2'=(\mathbf{x}_2',it_2)$. Setting $t_1=0$ and $t_2=0$ in the ground-state BS wave function, we obtain a description for the prepared state (ps)
\begin{equation}
\begin{split}\label{BSWFT0}
\mathscr{X}^{\text{ps}}_a=\chi_{P}(\mathbf{x}_1',t_1=0,\mathbf{x}_2',t_2=0)=\frac{1}{(2\pi)^{3/2}}\frac{1}{\sqrt{2E(P)}}e^{i\mathbf{P}\cdot(\eta_1\mathbf{x}_1'+\eta_2\mathbf{x}_2')}\chi_P(\mathbf{x}_1'-\mathbf{x}_2'),
\end{split}
\end{equation}
where $E(p)=\sqrt{\mathbf{p}^2+m^2}$ and $\eta_1+\eta_2=1$. We emphasize that the prepared state is not the physical state and the prepared state mass $M_0$ is not the physical mass of resonance.

Now it is necessary to consider the total Hamiltonian
\begin{equation}
\begin{split}
H=K_I+V_I,
\end{split}
\end{equation}
where $K_I$ represents the interaction responsible for the formation of stationary bound state and $V_I$ stands for the interaction responsible for the decay of resonance. Then the time evolution of this system determined by the total Hamiltonian $H$ has the explicit form
\begin{equation}
\begin{split}\label{timeevo}
\mathscr{X}(t)=e^{-iHt}\mathscr{X}^{\text{ps}}_{a}=\frac{1}{2\pi i}\int_{C_2}d\epsilon e^{-i\epsilon t}\frac{1}{\epsilon-H}\mathscr{X}^{\text{ps}}_{a},
\end{split}
\end{equation}
where $(\epsilon-H)^{-1}$ is the Green's function and the contour $C_2$ runs from $ic_r+\infty$ to $ic_r-\infty$ in energy-plane. The positive constant $c_r$ is sufficiently large that no singularity of $(\epsilon-H)^{-1}$ lies above $C_2$. The time-dependent wave function $\mathscr{X}(t)$ provides a complete description of the system for $t>0$. Since $H\neq K_I$, this system should not remain in the prepared state $\mathscr{X}^{\text{ps}}_a$. Then at arbitrary time $t$ the probability amplitude of finding the system in the state $\mathscr{X}^{\text{ps}}_a$ is
\begin{equation}
\begin{split}
\mathscr{A}_a=(\mathscr{X}^{\text{ps}}_{a},\mathscr{X}(t))=\frac{1}{2\pi i}\int_{C_2}d\epsilon\frac{e^{-i\epsilon t}}{\epsilon-M_0-(2\pi)^3T_{aa}(\epsilon)}.
\end{split}
\end{equation}
In field theory the operator $T(\epsilon)$ is just the scattering matrix with energy $\epsilon$, and $T_{aa}(\epsilon)$ is the $T$-matrix element between two bound states, which is defined as
\begin{equation}
\begin{split}
\langle a~\text{out}|a~\text{in}\rangle=\langle a~\text{in}|a~\text{in}\rangle-i(2\pi)^4\delta^{(4)}(P-P)T_{aa}(\epsilon).
\end{split}
\end{equation}

Because of the analyticity of $T_{aa}(\epsilon)$, we define
\begin{equation}
\begin{split}\label{Taepsilon}
T_{aa}(\epsilon)=\mathbb{D}(\epsilon)-i\mathbb{I}(\epsilon),
\end{split}
\end{equation}
where $\epsilon$ approaches the real axis from above, $\mathbb{D}$ and $\mathbb{I}$ are the real and imaginary parts, respectively. Using the unitarity of $T_{aa}(\epsilon)$, we obtain \cite{GreenFun}
\begin{equation}
\begin{split}\label{Taepsilon1}
2\mathbb{I}(\epsilon)=\sum_{c'}\sum_b(2\pi)^4\delta^{(3)}(\mathbf{P}_b-\mathbf{P})\delta(E_b-\epsilon)|T_{(c';b)a}(\epsilon)|^2,
\end{split}
\end{equation}
where $P_b=(\mathbf{P}_b,iE_b)$ is the total energy-momentum vector of all particles in the final state and the $T$-matrix element $T_{(c';b)a}(\epsilon)$ is defined as $\langle (c';b)~\text{out}|a~\text{in}\rangle=-i(2\pi)^4\delta^{(3)}(\mathbf{P}_b-\mathbf{P})\delta(E_b-\epsilon)T_{(c';b)a}(\epsilon)$ in channel $c'$. The delta-function in Eq. (\ref{Taepsilon1}) means that the energy $\epsilon$ in scattering matrix is equal to the total energy $E_b$ of the final state, $\sum_b$ represents summing over momenta and spins of all particles in the final state, and $\sum_{c'}$ represents summing over all possible channels. For $E_b=\epsilon$, we also denote the total energy of the final state by $\epsilon$ and $\mathbb{I}(\epsilon)$ becomes a function of the final state energy. Using dispersion relation for the function $T_{aa}(\epsilon)$, we obtain
\begin{equation}
\begin{split}\label{disrel}
\mathbb{D}(\epsilon)=-\frac{\mathcal{P}}{\pi}\int_{\epsilon_M}^\infty \frac{\mathbb{I}(\epsilon')}{\epsilon'-\epsilon}d\epsilon',
\end{split}
\end{equation}
where the symbol $\mathcal{P}$ means that this integral is a principal value integral and the variable of integration is the total energy $\epsilon'$ of the final state. To calculate the real part, we need calculate the function $\mathbb{I}(\epsilon')$ of value of the final state energy $\epsilon'$, which is an arbitrary real number over the real interval $\epsilon_M<\epsilon'<\infty$. As usual the momentum of initial bound state $a$ is set as $P=(0,0,0,iM_0)$ in the rest frame and $\epsilon_M$ denotes the sum of all particle masses in the final state. We suppose that the final state $b$ may contain $n$ composite particles and $n'$ elementary particles in decay channel $c'$. From Eq. (\ref{Taepsilon1}), we have
\begin{equation}
\begin{split}\label{Iepsilon'}
\mathbb{I}(\epsilon')=&\frac{1}{2}\sum_{c'}\int d^3Q'_1\cdots d^3Q'_{n'}d^3Q_1\cdots d^3Q_n(2\pi)^4\delta^{(4)}(Q'_1+\cdots+Q_n-P^{\epsilon'})\sum_{\text{spins}}|T_{(c';b)a}(\epsilon')|^2,
\end{split}
\end{equation}
where $\sum_{c'}$ represents summing over all open and closed channels, $Q'_1\cdots Q'_{n'}$ and $Q_1\cdots Q_n$ are the momenta of final elementary and composite particles, respectively; $P^{\epsilon'}=(0,0,0,i\epsilon')$, $T_{(c';b)a}(\epsilon')$ is the $T$-matrix element with respect to $\epsilon'$, and $\sum_{\text{spins}}$ represents summing over spins of all particles in the final state. In Eq. (\ref{Iepsilon'}) the energy in scattering matrix is equal to the total energy $\epsilon'$ of the final state $b$, which is an arbitrary real number over the real interval $\epsilon_M<\epsilon'<\infty$. Because the total energy $\epsilon'$ of the final state extends from $\epsilon_M$ to $+\infty$, we may obtain several closed channels derived from the interaction Lagrangian. The mass $M_0$ and BS amplitude of initial bound state $a$ have been specified and the value of the initial state energy in the rest frame is a specified value $M_0$. From Eq. (\ref{Iepsilon'}), we have $\mathbb{I}(\epsilon')>0$ for $\epsilon'>\epsilon_M$ and $\mathbb{I}(\epsilon')=0$ for $\epsilon'\leqslant\epsilon_M$, which is the reason that the integration in dispersion relation (\ref{disrel}) ranges from $\epsilon_M$ to $+\infty$.

In experiments, many exotic particles are narrow states and their decay widths are very small compared with their energy levels, i.e., $(2\pi)^3\mathbb{I}(M_0)\ll M_0$. This situation is ordinarily interpreted as implying that both $(2\pi)^3|\mathbb{D}(\epsilon)|$ and $(2\pi)^3\mathbb{I}(\epsilon)$ are also very small quantities, as compared to $M_0$. Therefore, we can expect that $[\epsilon-M_0-(2\pi)^3T_{aa}(\epsilon)]^{-1}$ has a pole on the second Riemann sheet
\begin{equation}
\begin{split}\label{pole}
\epsilon_{\text{pole}}\cong M_0+(2\pi)^3[\mathbb{D}(M_0)-i\mathbb{I}(M_0)]=M-i\frac{\Gamma(M_0)}{2},
\end{split}
\end{equation}
where $\Delta M=(2\pi)^3\mathbb{D}(M_0)$ is the correction for energy level of resonance and $M=M_0+(2\pi)^3\mathbb{D}(M_0)$ is the physical mass for resonance. This pole at $\epsilon_{\text{pole}}$ describes the resonance. The mass $M_0$ of two-body bound state is obtained by solving homogeneous BS equation, which should not be the mass of physical resonance. $\Gamma(M_0)$ with mass $M_0$ also should not be the width of physical resonance, which should depend on its physical mass $M$.

\section{BS equation for bound state composed of two vector mesons}\label{sec:BS}
At low energy QCD, an important feature is the spontaneous breaking of chiral symmetry, and the quark-gluon coupling constant becomes very large. Based on the spontaneous breaking of chiral symmetry, the effective theory at low energy QCD has been constructed \cite{QTFII}. According to the effective theory at low energy QCD, the vacuum state is not unique and non-vanishing vacuum quark condensate causes the spontaneous breaking of chiral symmetry, which leads to the appearance of Goldstone bosons. One can define the constituent quark fields, from which the Goldstone mode has been eliminated. The current quark fields can be written as the product of Goldstone boson fields and constituent quark fields. At low energy QCD, the effective interaction Lagrangian can be regarded as Lagrangian for the interaction of light mesons with constituent quarks. We emphasize that the constituent quark fields and light meson fields are the basic fields which appear in the effective interaction Lagrangian. In this paper, we investigate the light meson interaction with the light quarks in heavy mesons and the interaction Lagrangian for the coupling of constituent quark fields to light meson fields is \cite{mypaper6}
\begin{equation}\label{Lag}
\begin{split}
&\mathscr{L}^{\text{eff}}_I=ig_0\left(\begin{array}{ccc} \bar{u}&\bar{d}&\bar{s}
\end{array}\right)\gamma_5\left(\begin{array}{ccc}
\pi^0+\frac{1}{\sqrt{3}}\eta &\sqrt{2}\pi^+&\sqrt{2}K^+\\\sqrt{2}\pi^-&-\pi^0+\frac{1}{\sqrt{3}}\eta &\sqrt{2}K^0\\\sqrt{2}K^-&\sqrt{2}\bar{K}^0&-\frac{2}{\sqrt{3}}\eta
\end{array}\right)\left(\begin{array}{c} u\\d\\s
\end{array}\right)\\
&+ig'_0\left(\begin{array}{ccc} \bar{u}&\bar{d}&\bar{s}
\end{array}\right)\gamma_\mu\left(\begin{array}{ccc}
\rho^0+\omega&\sqrt{2}\rho^+&\sqrt{2}K^{*+}\\\sqrt{2}\rho^-&-\rho^0+\omega&\sqrt{2}K^{*0}\\\sqrt{2}K^{*-}&\sqrt{2}\bar{K}^{*0}&\sqrt{2}\phi
\end{array}\right)_\mu\left(\begin{array}{c} u\\d\\s
\end{array}\right)+g_\sigma\left(\begin{array}{cc} \bar{u}&\bar{d}
\end{array}\right)\left(\begin{array}{c} u\\d
\end{array}\right)\sigma.
\end{split}
\end{equation}
From this effective interaction Lagrangian, we have to consider that heavy meson is a bound state composed of a quark and an antiquark and investigate the interaction of light meson with constituent quarks in heavy meson. Besides, at low energy QCD we use the effective interaction Lagrangian for the coupling of constituent quark fields to light meson fields, and we employ perturbative QCD to deal with the interaction involving heavy quark, which is consistent with QCD.

In this paper, we assume that resonance $T_{c\bar c}(4020)^+$ is an unstable molecular state $D^{*+}\bar{D}^{*0}$ with spin-parity quantum numbers $1^+$. As mentioned in Sec. \ref{sec:DBST}, there are two steps to deal with this unstable system. As the first step, we investigate the stable bound state $D^{*+}\bar{D}^{*0}$, which is considered as the prepared state. As the second step, we study the time evolution of unstable system determined by the total Hamiltonian and obtain the correction for energy level of resonance due to decay channels. In this section, our attention is only focused on the bound state composed of two vector mesons.

If a bound state with spin $j$ and parity $\eta_{P}$ is created by two Heisenberg vector fields with masses $M_1$ and $M_2$, respectively, its BS wave function is defined as
\begin{equation}\label{BSwfdx}
\chi_{P(\lambda\tau)}^j(x_1',x_2')=\langle0|TA_\lambda(x_1')A^\dagger_\tau(x_2')|P,j\rangle=\frac{1}{(2\pi)^{3/2}}\frac{1}{\sqrt{2E(P)}}e^{iP\cdot X}\chi_{P(\lambda\tau)}^j(X'),
\end{equation}
where $P$ is the momentum of the bound state, $X=\eta_1x_1'+\eta_2x_2'$, $X'=x_1'-x_2'$ and $\eta_{1,2}=M_{1,2}/(M_1+M_2)$. Making the Fourier transformation, we obtain BS wave function in the momentum representation
\begin{equation}\label{BSwfdp}
\begin{split}
\chi^j_P(p_1',p_2')_{\lambda\tau}=\frac{1}{(2\pi)^{3/2}}\frac{1}{\sqrt{2E(P)}}(2\pi)^4\delta^{(4)}(P-p_1'+p_2')\chi^j_{\lambda\tau}(P,p),
\end{split}
\end{equation}
where $p$ is the relative momentum of two vector fields and we have $P=p_1'-p_2'$, $p=\eta_2p_1'+\eta_1p_2'$, $p_1'$ and $p_2'$ are the momenta carried by two vector fields, respectively. In Ref. \cite{mypaper9}, we have given the general form of BS wave functions for the bound states created by two massive vector fields with arbitrary spin and definite parity, for $\eta_{P}=(-1)^j$,
\begin{equation}\label{jp}
\begin{split}
\chi_{\lambda\tau}^{j}(P,p)=&\frac{1}{\mathcal{N}^j}\eta_{\mu_1\cdots\mu_j}[p_{\mu_1}\cdots p_{\mu_j}(\mathcal{T}^1_{\lambda\tau}\Phi_1+\mathcal{T}^2_{\lambda\tau}\Phi_2)+\mathcal{T}^3_{\mu_1\cdots\mu_j\lambda\tau}\Phi_3+\mathcal{T}^4_{\mu_1\cdots\mu_j\lambda\tau}\Phi_4\\
&+\mathcal{T}^5_{\mu_1\cdots\mu_j\lambda\tau}\Phi_5+\mathcal{T}^6_{\mu_1\cdots\mu_j\lambda\tau}\Phi_6],
\end{split}
\end{equation}
for $\eta_{P}=(-1)^{j+1}$,
\begin{equation}\label{jm}
\begin{split}
\chi_{\lambda\tau}^{j}(P,p)=&\frac{1}{\mathcal{N}^j}\eta_{\mu_1\cdots\mu_j}(p_{\mu_1}\cdots p_{\mu_j}\epsilon_{\lambda\tau\xi\zeta}p_\xi P_\zeta\Phi'_1+\mathcal{T}^7_{\mu_1\cdots\mu_j\lambda\tau}\Phi'_2+\mathcal{T}^8_{\mu_1\cdots\mu_j\lambda\tau}\Phi'_3+\mathcal{T}^{9}_{\mu_1\cdots\mu_j\lambda\tau}\Phi'_4\\
&+\mathcal{T}^{10}_{\mu_1\cdots\mu_j\lambda\tau}\Phi'_5+\mathcal{T}^{11}_{\mu_1\cdots\mu_j\lambda\tau}\Phi'_6+\mathcal{T}^{12}_{\mu_1\cdots\mu_j\lambda\tau}\Phi'_7),
\end{split}
\end{equation}
where $\mathcal{N}^j$ is normalization, $\eta_{\mu_1\cdots\mu_j}$ is the polarization tensor describing the spin of the bound state, subscripts $\lambda$ and $\tau$ are derived from these two vector fields, the independent tensor structures $\mathcal{T}^i_{\lambda\tau}$ are given in Appendix \ref{app1}, $\Phi_i(P\cdot p,p^2)$ and $\Phi'_i(P\cdot p,p^2)$ are independent scalar functions.

From Eq. (\ref{jm}), we obtain BS wave function describing bound state $D^{*+}\bar{D}^{*0}$ with spin-parity quantum numbers $1^+$
\begin{equation}\label{BSWF1+}
\begin{split}
\chi_{\lambda\tau}^{1^+}(P,p)=&\frac{1}{\mathcal{N}^{1^+}}\eta^{\varsigma=1,2,3}_{\mu}(P)\chi_{\mu\lambda\tau}(P,p)\\
=&\frac{1}{\mathcal{N}^{1^+}}\eta^{\varsigma=1,2,3}_{\mu}(P)(p_{\mu}\epsilon_{\lambda\tau\xi\zeta}p_\xi P_\zeta\mathcal{G}_1+\mathcal{T}^7_{\mu\lambda\tau}\mathcal{G}_2+\mathcal{T}^8_{\mu\lambda\tau}\mathcal{G}_3+\mathcal{T}^{9}_{\mu\lambda\tau}\mathcal{G}_4+\mathcal{T}^{10}_{\mu\lambda\tau}\mathcal{G}_5).
\end{split}
\end{equation}
There are five independent components in Eq. (\ref{BSWF1+}). This BS wave function should satisfy BS equation
\begin{equation}\label{BSE}
\chi^{1^+}_{\lambda\tau}(P,p)=-\int \frac{d^4p'}{(2\pi)^4}\Delta_{F\lambda\theta}(p_1')\mathcal{V}_{\theta\theta',\kappa'\kappa}(p,p';P)\chi^{1^+}_{\theta'\kappa'}(P,p')\Delta_{F\kappa\tau}(p_2'),
\end{equation}
where $\mathcal{V}_{\theta\theta',\kappa'\kappa}$ is the interaction kernel, $P=(0,0,0,iM_0)$, $p_1'=p+P/2$, $p_2'=p-P/2$, $\Delta_{F\lambda\theta}(p_1')$ and $\Delta_{F\kappa\tau}(p_2')$ are the propagators for the spin 1 fields, $\Delta_{F\lambda\theta}(p_1')=(\delta_{\lambda\theta}+\frac{p'_{1\lambda} p'_{1\theta}}{M_1^2})\frac{-i}{p_1'^2+M_1^2-i\varepsilon}$, $\Delta_{F\kappa\tau}(p_2')=(\delta_{\kappa\tau}+\frac{p'_{2\kappa} p'_{2\tau}}{M_2^2})\frac{-i}{p_2'^2+M_2^2-i\varepsilon}$, $M_1=M_{D^{*+}}$ and $M_2=M_{\bar{D}^{*0}}$. We emphasize that the kernel $\mathcal{V}$ is defined in two-body channel so $\mathcal{V}$ is not complete interaction. The kernel in homogeneous BS equation (\ref{BSE}) plays a central role for making two-body system to be a stable bound state, and the solution of homogeneous BS equation (\ref{BSE}) should only describe bound state. In our approach, BS equation for meson-meson bound state is treated in the ladder approximation. This approximation consists in replacing the interaction kernel by its lowest order value corresponding to the simple one-meson exchange.

\begin{figure}[tbp]
\centering 
\includegraphics[width=.49\textwidth,clip]{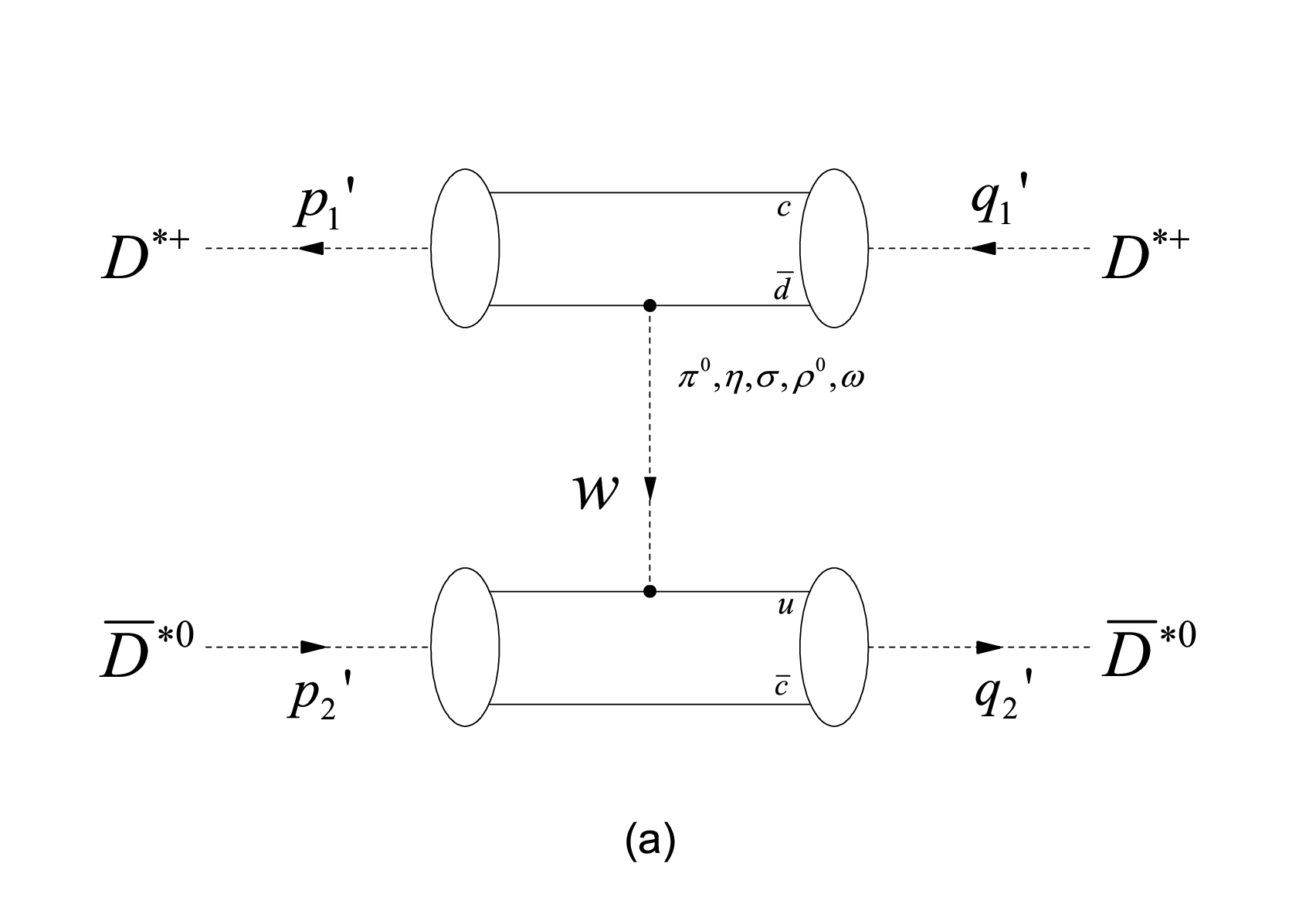}
\hfill
\includegraphics[width=.49\textwidth,origin=c]{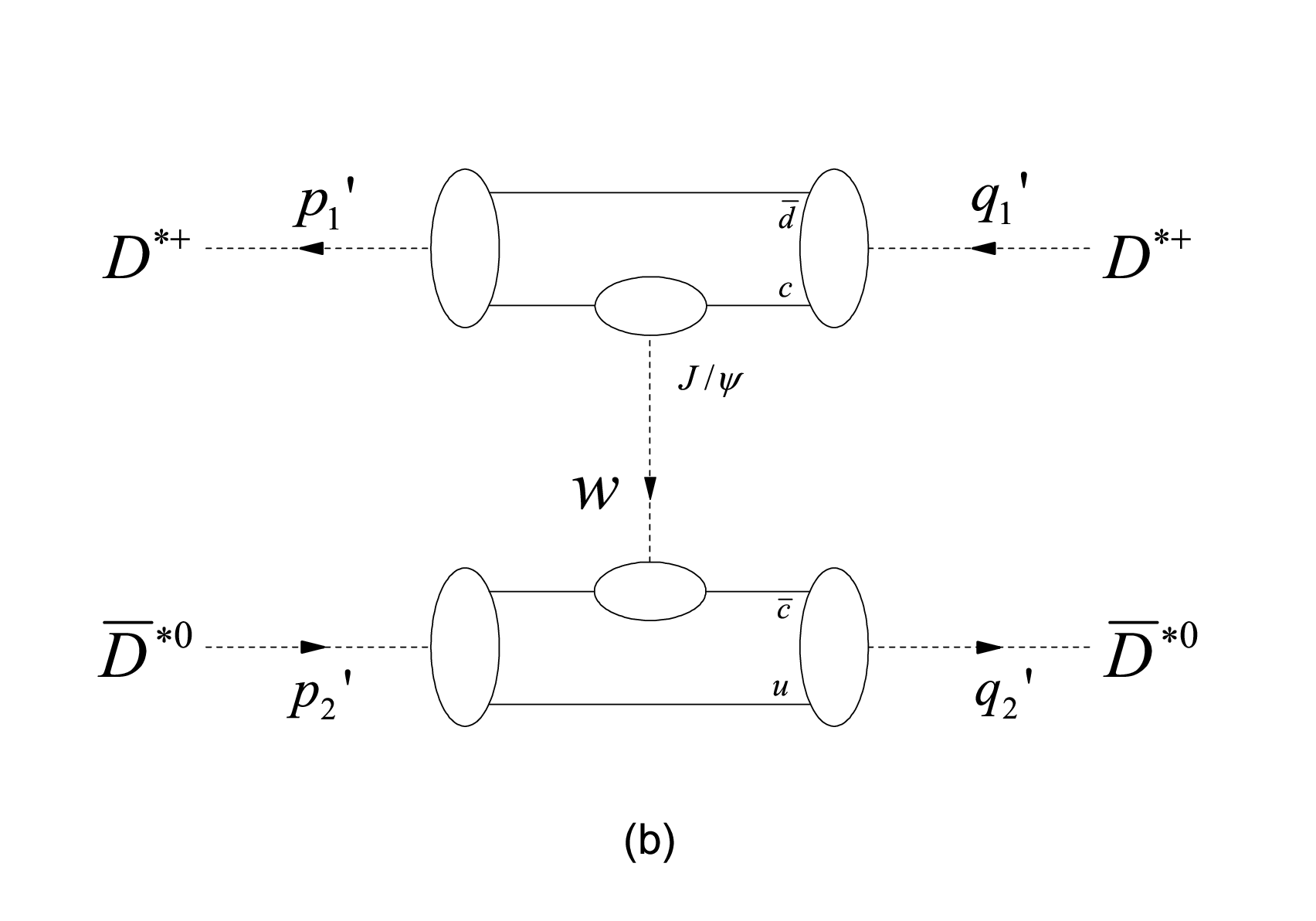}
\caption{\label{Fig1} One meson exchange between two quarks in two heavy vector mesons. The solid lines denote quark propagators, and the unfilled ellipses represent Bethe-Salpeter amplitudes. In diagram (a) the light meson interaction with the light quarks in heavy meson is shown; in diagram (b) the heavy vector meson $J/\psi$ is considered as a bound state of $c\bar c$.}
\end{figure}

Different from the previous works about hadronic molecules \cite{ms:Guo,ms:Dong,ms:Liu}, in our approach the heavy mesons in a molecular state are considered as bound states composed of a heavy quark and a light quark. From the interaction Lagrangian for the coupling of light quark fields to light meson fields expressed as Eq. (\ref{Lag}), we can obtain the interaction kernel between two light quarks in two heavy mesons from one light meson exchange. Moreover, we should consider the interaction kernel between two heavy quarks in two heavy mesons from one heavy meson exchange. In our theoretical frame, the interaction kernel between two heavy mesons is derived from one meson exchange between two quarks in these two heavy mesons. To construct the interaction kernel between $D^{*+}$ and $\bar D^{*0}$, we consider one light meson ($\pi^0$, $\eta$, $\sigma$, $\rho^0$, $\omega$) exchange and one heavy meson ($J/\psi$) exchange. The light meson exchange between two light quarks in two heavy vector mesons is shown in Fig. \ref{Fig1}a. The heavy meson exchange between two heavy quarks in two heavy vector mesons is represented by the graph of Fig. \ref{Fig1}b. The heavy vector meson $J/\psi$ is considered as a bound state of $c\bar c$, and in Fig. \ref{Fig1}b BS amplitude of $J/\psi$ is represented by the unfilled ellipse. Using the approach introduced in Refs. \cite{mypaper3,mypaper4,mypaper5,mypaper9}, we can obtain the interaction kernel from one light meson ($\pi^0$, $\eta$, $\sigma$, $\rho^0$, $\omega$) exchange and one heavy meson ($J/\psi$) exchange
\begin{equation}\label{kernel1}
\begin{split}
\mathcal {V}&_{\theta\theta',\kappa'\kappa}(p,p';P)=h^{(\text{p})}(w^{2})\epsilon_{\iota\iota'\theta\theta'}p_{1\iota}'q_{1\iota'}'\bigg(\frac{ig_\pi^2}{w^2+M_\pi^2}+\frac{-ig_\eta^2/3}{w^2+M_\eta^2}\bigg)\bar h^{(\text{p})}(w^{2})\epsilon_{\varpi\varpi'\kappa'\kappa}p_{2\varpi}'q_{2\varpi'}'\\
&+h_1^{(\text{s})}(w^2)\frac{-ig_\sigma^2}{w^2+M_\sigma^2}\bar{h}_1^{(\text{s})}(w^2)\delta_{\theta\theta'}\delta_{\kappa'\kappa}+\bigg(\frac{ig_\rho^2}{w^2+M_\rho^2}+\frac{-ig_\omega^2}{w^2+M_\omega^2}\bigg)\{h_1^{(\text{lv})}(w^2)\bar{h}_1^{(\text{lv})}(w^2)\\
&\times(p_1'+q_1')\cdot(-p_2'-q_2')\delta_{\theta\theta'}\delta_{\kappa'\kappa}+h_1^{(\text{lv})}(w^2)\bar{h}_2^{(\text{lv})}(w^2)\delta_{\theta\theta'}[(p_1'+q_1')_{\kappa'} q_{2\kappa}'+p_{2\kappa'}'(p_1'+q_1')_\kappa]\\
&+h_2^{(\text{lv})}(w^2)\bar{h}_1^{(\text{lv})}(w^2)[q_{1\theta}'(p_2'+q_2')_{\theta'}+(p_2'+q_2')_\theta p_{1\theta'}']\delta_{\kappa'\kappa}-h_2^{(\text{lv})}(w^2)\bar{h}_2^{(\text{lv})}(w^2)[q_{1\theta}'\delta_{\theta'\kappa'}q_{2\kappa}'\\
&+q_{1\theta}'\delta_{\theta'\kappa}p_{2\kappa'}'+\delta_{\theta\kappa'}p_{1\theta'}'q_{2\kappa}'+\delta_{\theta\kappa}p_{1\theta'}'p_{2\kappa'}']\}+\frac{-i}{w^2+M_{J/\psi}^2}\{h_1^{(\text{hv})}(w^2)\bar{h}_1^{(\text{hv})}(w^2)\\
&\times(p_1'+q_1')\cdot(-p_2'-q_2')\delta_{\theta\theta'}\delta_{\kappa'\kappa}+h_1^{(\text{hv})}(w^2)\bar{h}_2^{(\text{hv})}(w^2)\delta_{\theta\theta'}[(p_1'+q_1')_{\kappa'}q_{2\kappa}'+p_{2\kappa'}'(p_1'+q_1')_\kappa]\\
&+h_2^{(\text{hv})}(w^2)\bar{h}_1^{(\text{hv})}(w^2)[q_{1\theta}'(p_2'+q_2')_{\theta'}+(p_2'+q_2')_\theta p_{1\theta'}']\delta_{\kappa'\kappa}-h_2^{(\text{hv})}(w^2)\bar{h}_2^{(\text{hv})}(w^2)[q_{1\theta}'\delta_{\theta'\kappa'}q_{2\kappa}'\\
&+q_{1\theta}'\delta_{\theta'\kappa}p_{2\kappa'}'+\delta_{\theta\kappa'}p_{1\theta'}'q_{2\kappa}'+\delta_{\theta\kappa}p_{1\theta'}'p_{2\kappa'}']\},
\end{split}
\end{equation}
where $g$ represents the corresponding meson-quark coupling constant, $g_\pi=\frac{340}{93}$, $g_\sigma=\frac{B(M_\sigma)}{f_\sigma}=\frac{307}{60}$ \cite{cc1a,cc1b}, $g_\omega^2=g_\rho^2=2.42/2$ \cite{cc2}; $p'_1=(\mathbf{p},ip_{10}')$, $p'_2=(\mathbf{p},ip_{20}')$, $q'_1=(\mathbf{p}',iq_{10}')$, $q'_2=(\mathbf{p}',iq_{20}')$, $w=q_1'-p_1'=q_2'-p_2'$ is the momentum of the exchanged meson and $\mathbf{w}=\mathbf{p}'-\mathbf{p}$; $h(w^2)$ and $\bar h(w^2)$ are scalar functions. The details in computational process are shown in Appendix \ref{app2}. Then we can solve BS equation (\ref{BSE}) and the mass $M_0$ and BS wave function of bound state $D^{*+}\bar{D}^{*0}$ with spin-parity quantum numbers $1^+$ can be obtained. The procedure for solving BS equation is shown in Appendix \ref{app3}. The reduced normalization condition for $\chi_{\lambda\tau}^{1^+}(P,p)$ expressed as Eq. (\ref{BSWF1+}) is
\begin{equation}
\begin{split}
&\frac{1}{3}\sum_{\varsigma=1}^3\frac{-i}{(2\pi)^4}\int d^4p\bar\chi_{\lambda'\tau'}(P,p)\frac{\partial}{\partial P_0}[\Delta_{F\lambda'\lambda}(p+P/2)^{-1}\Delta_{F\tau\tau'}(p-P/2)^{-1}]\chi_{\lambda\tau}(P,p)=(2P_0)^2,
\end{split}
\end{equation}
where $\Delta_{F\beta\alpha'}(p)^{-1}$ is the inverse propagator for the vector field with mass $m$, $\Delta_{F\beta\alpha'}(p)^{-1}=i(\delta_{\beta\alpha'}-\frac{p_{\beta}p_{\alpha'}}{p^2+m^2})(p^2+m^2)$ \cite{mypaper6}.

The heavy meson is a bound state consisting of a quark and an antiquark and the meson-meson bound state is actually a four-quark state. The generalized BS wave function for four-quark state describing the bound state composed of $D^{*+}\bar{D}^{*0}$ with spin-parity quantum numbers $1^+$ has the form \cite{fourbody1,mypaper6,mypaper7}
\begin{equation}\label{fourquarkBSWF1}
\begin{split}
\chi^{1^+}(P,p,k,k')=\chi_{\lambda}(p_1',k)\chi^{1^+}_{\lambda\tau}(P,p)\chi_\tau(p_2',k'),
\end{split}
\end{equation}
where $\chi_{\lambda}(p_1',k)$ and $\chi_\tau(p_2',k')$ are BS wave functions of $D^{*+}$ and $\bar{D}^{*0}$ mesons, respectively; $k$ and $k'$ are the relative momenta between quark and antiquark in these two mesons. BS wave function of $D^{*}$ meson will be given in Sec. \ref{sec:hcpie'}. Then we obtain BS wave function of the prepared state. It is necessary to emphasize that the prepared state is not the physical state and the prepared state mass $M_0$ ($M_0<M_{D^{*+}}+M_{\bar D^{*0}}$) is not the physical mass of resonance.

\section{$T$-matrix element $T_{(c';b)a}(\epsilon')$}\label{sec:T-ME}
In order to obtain the real part $\mathbb{D}(\epsilon)$ due to all decay channels, we have to calculate $T_{(c';b)a}(\epsilon')$ with respect to value of the final state energy $\epsilon'$, which is an arbitrary real number over the real interval $\epsilon_M<\epsilon'<\infty$. In experiments \cite{X40201,X40202,X40203,X40204,X40205} two strong decay modes of $T_{c\bar c}(4020)^+$ have been observed: $h_c(1P)\pi^+$ denoted as $c'_1$ and $D^{*+}\bar D^{*0}$ denoted as $c'_2$. We emphatically introduce the $T$-matrix elements $T_{(c'_1;b)a}(\epsilon')$ and $T_{(c'_2;b)a}(\epsilon')$ in this section.

\subsection{Channel $h_c(1P)\pi^+$}\label{sec:hcpie'}
In our approach, the heavy vector meson $D^*$ is a bound state consisting of a quark and an antiquark and the initial meson-meson bound state is actually composed of four quarks. Here, we consider that in the final state the light pseudoscalar meson $\pi$ is an elementary particle and the heavy axial-vector meson $h_c(1P)$ is a bound state of $c\bar c$. Mandelstam's approach is a technique based on BS wave function for evaluating the general matrix element between bound states. Applying Mandelstam's approach, one can express the general matrix element between bound states in terms of BS wave functions and a two-particle irreducible Green's function \cite{ParticlesFields}. In DBST \cite{mypaper8}, Mandelstam's approach has been generalized to evaluate the bound state matrix element with respect to arbitrary value of the final state energy $\epsilon'$.

For heavy vector mesons, the authors of Refs. \cite{BSE:Roberts1,BSE:Roberts3,BSE:Roberts4,BSE:Roberts5} have obtained their BS amplitudes in Euclidean space:
\begin{equation}\label{BSA}
\begin{split}
\Gamma_{\lambda}^{V}(K,k)=\frac{1}{\mathcal{N}^{V}}\bigg(\gamma_\lambda+K_{\lambda}\frac{\gamma\cdot K}{M_{V}^2}\bigg)\varphi_{V}(k^2),
\end{split}
\end{equation}
where $K$ is the momentum of heavy meson, $k$ denotes the relative momentum between quark and antiquark in heavy meson, $M_V$ is heavy vector meson mass, $\Gamma_{\lambda}^{V}(K,k)$ is transverse ($K_{\lambda}\Gamma_{\lambda}^{V}(K,k)=0$), $\mathcal{N}^V$ is normalization, and $\varphi_{V}(k^{2})$ is scalar function fixed by providing fits to observables. The charmed meson $D^{*+}$ is composed of $c$-quark and $d$-antiquark. As in heavy-quark effective theory (HQET) \cite{hqet}, we consider that the heaviest quark carries all the heavy-meson momentum and obtain BS wave function of $D^{*+}$ meson
\begin{equation}\label{D0BSwf}
\begin{split}
\chi_\lambda(K,k)=\frac{-1}{\gamma\cdot(k+K)-im_c}\frac{1}{\mathcal{N}^{D^{*+}}}\bigg(\gamma_\lambda+K_{\lambda}\frac{\gamma\cdot K}{M_{D^{*+}}^2}\bigg)\varphi_{D^{*+}}(k^2)\frac{-1}{\gamma\cdot k-im_d},
\end{split}
\end{equation}
where $K$ is the momentum of heavy meson, $k$ becomes the relative momentum between $c$-quark and $d$-antiquark, $m_{c}$ is the heavy quark mass, $m_{d}$ is the constituent quark mass, $\varphi_{D^{*+}}(k^{2})=\varphi_{D^{*0}}(k^{2})=\exp(-k^{2}/\omega_{D^{*}}^{2})$ and $\omega_{D^{*}}$=1.50 GeV \cite{BSE:Roberts5}. The reduced normalization condition for the BS wave function of $D^{*+}$ meson given by Eq. (\ref{D0BSwf}) is
\begin{equation}
\begin{split}
\frac{-i}{(2\pi)^4}\frac{1}{3}\int d^4k\bar\chi_{\lambda}(K,k)\frac{\partial}{\partial K_0}[S_F(k+K)^{-1}]S_F(k)^{-1}\chi_\lambda(K,k)=(2K_0)^2,
\end{split}
\end{equation}
where $S_F(p)^{-1}$ is the inverse propagator for quark field and the factor $1/3$ appears because of the sum of three transverse directions.

For the heavy axial-vector meson $h_c(1P)$, we should find BS wave function for $P$-wave state of $c\bar c$. The authors of Ref. \cite{BSWF:Dai} have given the most general form of Bethe-Salpeter wave functions for mesons of arbitrary spin and definite parity. Considering the $P$-wave state, we obtain BS wave function of $h_c(1P)$ meson in Euclidean space
\begin{equation}\label{hcBSWF}
\begin{split}
\chi_{5\nu}(Q,q)&=\frac{1}{\mathcal{N}^{h_c}}\bigg[\gamma_5\bigg(q_\nu+Q_\nu\frac{Q\cdot q}{M_{h_c}^2}\bigg)\varphi_{h_c1}(Q\cdot q,q^2)+i\gamma_5\bigg(q_\nu+Q_\nu\frac{Q\cdot q}{M_{h_c}^2}\bigg)\gamma\cdot Q\varphi_{h_c2}(Q\cdot q,q^2)\bigg],
\end{split}
\end{equation}
where $Q$ is the momentum of heavy meson, $q$ becomes the relative momentum between $c$-quark and $c$-antiquark, $\chi_{5\nu}(Q,q)$ is transverse ($Q_{\nu}\chi_{5\nu}(Q,q)=0$), $\mathcal{N}^{h_c}$ is normalization, $\varphi_{h_c1}(Q\cdot q,q^2)$ and $\varphi_{h_c2}(Q\cdot q,q^2)$ are scalar functions. The heavy axial-vector meson $h_c(1P)$ with equal-mass constituents has quantum numbers $1^{+-}$, and scalar functions $\varphi_{h_c1}(Q\cdot q,q^2)$ and $\varphi_{h_c2}(Q\cdot q,q^2)$ in Eq. (\ref{hcBSWF}) should be even functions of $Q\cdot q$, as demonstrated in Ref. \cite{BSE:Krassnigg}. By fitting homogeneous Bethe-Salpeter equation for heavy axial-vector meson $h_c(1P)$, we obtain
\begin{equation}
\begin{split}
\varphi_{h_c1}(Q\cdot q,q^2)&=\frac{1}{(Q\cdot q)^2-\mathcal{C}_{\omega_1}}\exp\bigg(-\frac{q^2}{\omega^2_{h_c1}}\bigg),\\
\varphi_{h_c2}(Q\cdot q,q^2)&=\frac{1}{M_{h_c}}\frac{1}{(Q\cdot q)^2-\mathcal{C}_{\omega_2}}\exp\bigg(-\frac{q^2}{\omega^2_{h_c2}}\bigg),
\end{split}
\end{equation}
where $\mathcal{C}_{\omega_1}=(0.86~\text{GeV})^4$, $\omega_{h_c1}$=0.89 GeV, $\mathcal{C}_{\omega_2}=(0.88~\text{GeV})^4$ and $\omega_{h_c2}$=0.65 GeV. The procedure for determining BS wave function of heavy axial-vector meson $h_c(1P)$ is presented in Appendix \ref{app4}. The reduced normalization condition for BS wave function of $h_c(1P)$ meson expressed as Eq. (\ref{hcBSWF}) is
\begin{equation}
\begin{split}
\frac{-i}{(2\pi)^4}\frac{1}{3}\int d^4q\bar\chi_{5\nu}(Q,q)\frac{\partial}{\partial Q_0}[S_F(q+Q/2)^{-1}S_F(q-Q/2)^{-1}]\chi_{5\nu}(Q,q)=(2Q_0)^2,
\end{split}
\end{equation}
where the factor $1/3$ appears for the three transverse directions are summed.

\begin{figure}[!htb] \centering
\includegraphics[trim = 0mm 30mm 0mm 30mm,scale=1,width=10cm]{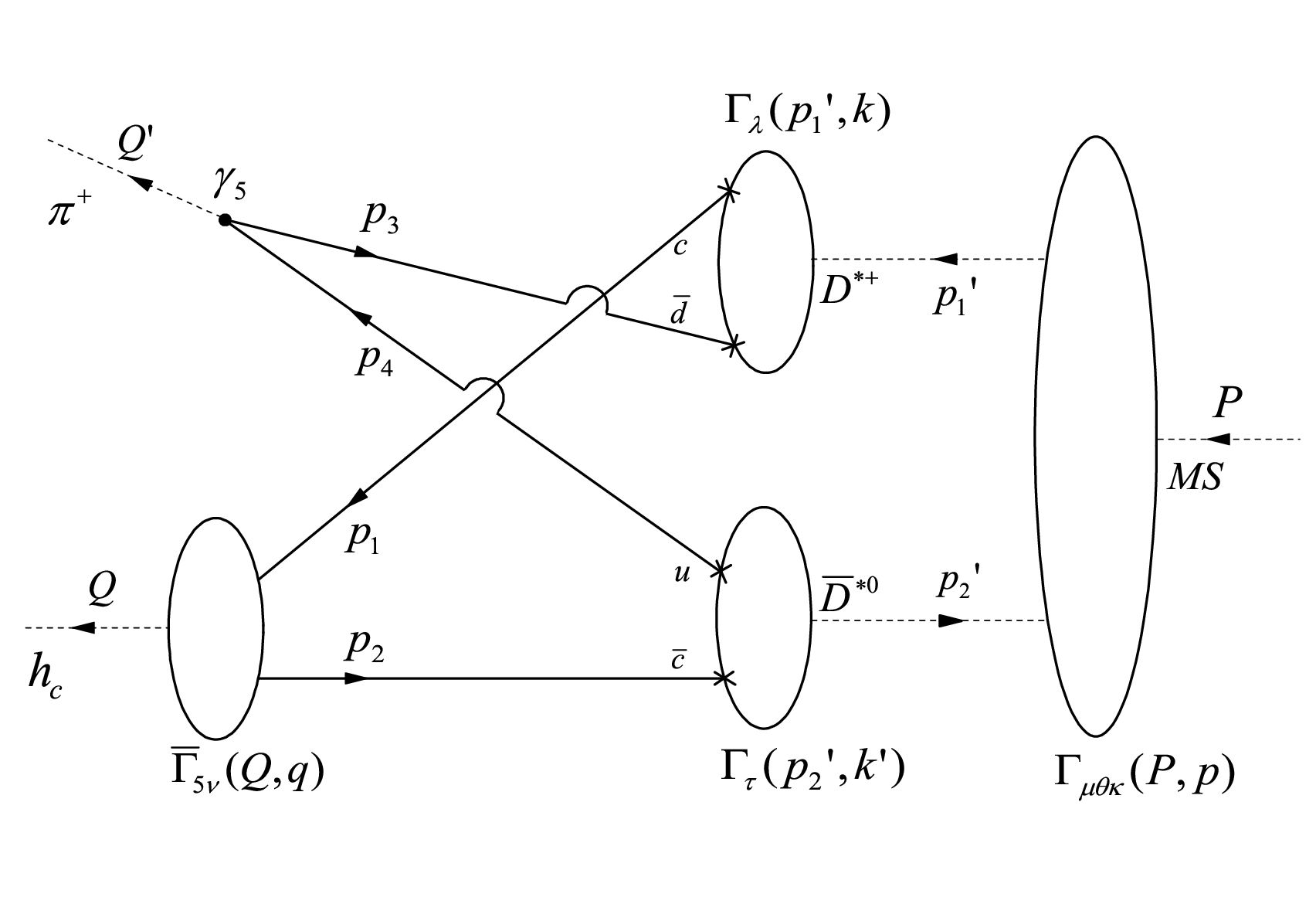}
\caption{\label{Fig2} Matrix element with respect to $\epsilon'$ for channel $h_c(1P)\pi^+$. The momenta in the final state satisfy $Q+Q'=P^{\epsilon'}$ and the momentum of the initial state is $P$. The final state energy extends from $\epsilon_{c'_1,M}$ to $+\infty$ while the initial state energy is specified, and the crosses mean that the momenta of quark propagators depend on the final state energy $\epsilon'$.}
\end{figure}

Applying the generalized Mandelstam's approach and retaining the lowest order term of the two-particle irreducible Green's function, we can obtain the $T$-matrix element $T_{(c'_1;b)a}(\epsilon')$ with respect to arbitrary value of the final state energy $\epsilon'$ for channel $c'_1$ in the momentum representation
\begin{equation}
\begin{split}\label{Tmatrele1}
T_{(c'_1;b)a}(\epsilon')=&\frac{-i}{(2\pi)^{9/2}}\frac{\sqrt2g_\pi}{\sqrt{2E_{\pi}(Q')}}\frac{\varepsilon_\nu^{\varrho}(Q)}{\sqrt{2E_{h_c}(Q)}}\frac{\eta^\varsigma_{\mu}(P)}{\sqrt{2E(P)}}\int \frac{d^4kd^4p}{(2\pi)^8}\\
&\times\text{Tr}[\bar\chi_{5\nu}(Q,q)\Gamma^{{D^{*+}}}_\lambda(p_1',k)S_F(p_3)\gamma_5 S_F(p_4)\Gamma^{{\bar D^{*0}}}_\tau(p_2',k')\frac{1}{\mathcal{N}^{1^+}}\chi_{\mu\lambda\tau}(P,p)],
\end{split}
\end{equation}
where the total energy $\epsilon'$ of the final state extends from $\epsilon_{c'_1,M}$ to $+\infty$, i.e., $\epsilon_{c'_1,M}<\epsilon'<\infty$; $Q$ and $Q'$ represent the momenta of final particles, $Q^2=-M_{h_c}^2$ and $Q'^2=-M_{\pi}^2$; $E(p)=\sqrt{\mathbf{p}^2+m^2}$; $p_1, p_3, p_4, p_2$ are the momenta of four quarks; $p_1'$ and $p_2'$ are the momenta of two heavy vector mesons; $q$, $k$ and $k'$ are the relative momenta between quark and antiquark in heavy mesons, respectively; $\varepsilon_\nu^{\varrho=1,2,3}(Q)$ and $\eta^{\varsigma=1,2,3}_{\mu}(P)$ are the polarization vectors of $h_c$ in the final state and bound state $D^{*+}\bar{D}^{*0}$ with spin-parity quantum numbers $1^+$ in the initial state, respectively; $\Gamma_\lambda^{D^{*}}(K,k)$ represents BS amplitude of heavy vector meson $D^{*}$, $S_F(p)$ is the quark propagator. BS wave function $\chi_{5\nu}(Q,q)$ for heavy axial-vector meson $h_c(1P)$ in the final state is expressed as Eq. (\ref{hcBSWF}), and BS wave function $\chi_{\mu\lambda\tau}(P,p)$ for initial meson-meson bound state ($MS$) is expressed as Eq. (\ref{BSWF1+}). It is necessary to emphasize that the energy in the two-particle irreducible Green's function is equal to the final state energy $\epsilon'$ while the mass $M_0$ and BS amplitude of initial bound state are specified. In our approach, the initial bound state is considered as a four-quark state \cite{mypaper6}, so the generalized BS amplitude of initial bound state should be $\Gamma^{D^{*+}}_\lambda(p_1',k)\chi_{\mu\lambda\tau}(P,p)\Gamma^{\bar{D}^{*0}}_\tau(p_2',k')$, which has been specified. We have introduced \emph{extended Feynman diagram} in Ref. \cite{mypaper8} to represent arbitrary value of the final state energy, shown as Fig. \ref{Fig2}. In Fig. \ref{Fig2}, the quark momenta in left-hand side of crosses depend on the final state energy and the momenta in right-hand side depend on the initial state energy, i.e., $p_1-p_2-p_3+p_4=Q+Q'=P^{\epsilon'}$ and $p_1'-p_2'=P$. In the rest frame, we have $P=(0,0,0,iM_0)$, $P^{\epsilon'}=(0,0,0,i\epsilon')$, $\epsilon_{c'_1,M}<\epsilon'<\infty$ and $\epsilon_{c'_1,M}=M_{h_c}+M_\pi$. When $\epsilon'=M_0$, the crosses in Fig. \ref{Fig2} disappear and then the extended Feynman diagram becomes the traditional Feynman diagram. These momenta should become
\begin{equation}\label{momenta}
\begin{split}
&p_1=(Q+Q')/2+p+k,~~p_2=(Q+Q')/2-Q+p+k,~~p_3=k,~~p_4=Q'+k,\\
&q=Q'/2+p+k,~k'=Q'(M_0)+k,~~p_1'=p+P/2,~~p_2'=p-P/2,~~Q+Q'=P^{\epsilon'},
\end{split}
\end{equation}
where $Q'(M_0)=(-\mathbf{Q}(M_0),i\sqrt{\mathbf{Q}^2(M_0)+M_\pi^2})$, $Q=(\mathbf{Q}(\epsilon'),i\sqrt{\mathbf{Q}^2(\epsilon')+M_{h_c}^2})$, $Q'=(-\mathbf{Q}(\epsilon'),i\sqrt{\mathbf{Q}^2(\epsilon')+M_\pi^2})$, $\mathbf{Q}^2(M_0)=[M_0^2-(M_{h_c}+M_\pi)^2][M_0^2-(M_{h_c}-M_\pi)^2]/(4M_0^2)$ and $\mathbf{Q}^2(\epsilon')=[\epsilon'^2-(M_{h_c}+M_\pi)^2][\epsilon'^2-(M_{h_c}-M_\pi)^2]/(4\epsilon'^2)$. From Eq. (\ref{Iepsilon'}), we obtain the function $\mathbb{I}_1(\epsilon')$ for channel $h_c(1P)\pi^+$
\begin{equation}\label{Iepsilon'1}
\begin{split}
\mathbb{I}_1(\epsilon')&=\frac{1}{2}\int d^3Qd^3Q'(2\pi)^4\delta^{(4)}(Q+Q'-P^{\epsilon'})\frac{1}{3}\sum_{\varsigma=1}^3\sum_{\varrho=1}^3|T_{(c'_1;b)a}(\epsilon')|^2.
\end{split}
\end{equation}

\subsection{Channel $D^{*+}\bar D^{*0}$}\label{sec:DDBARe'}
\begin{figure}[!htb] \centering
\includegraphics[trim = 0mm 50mm 0mm 40mm,scale=1,width=10cm]{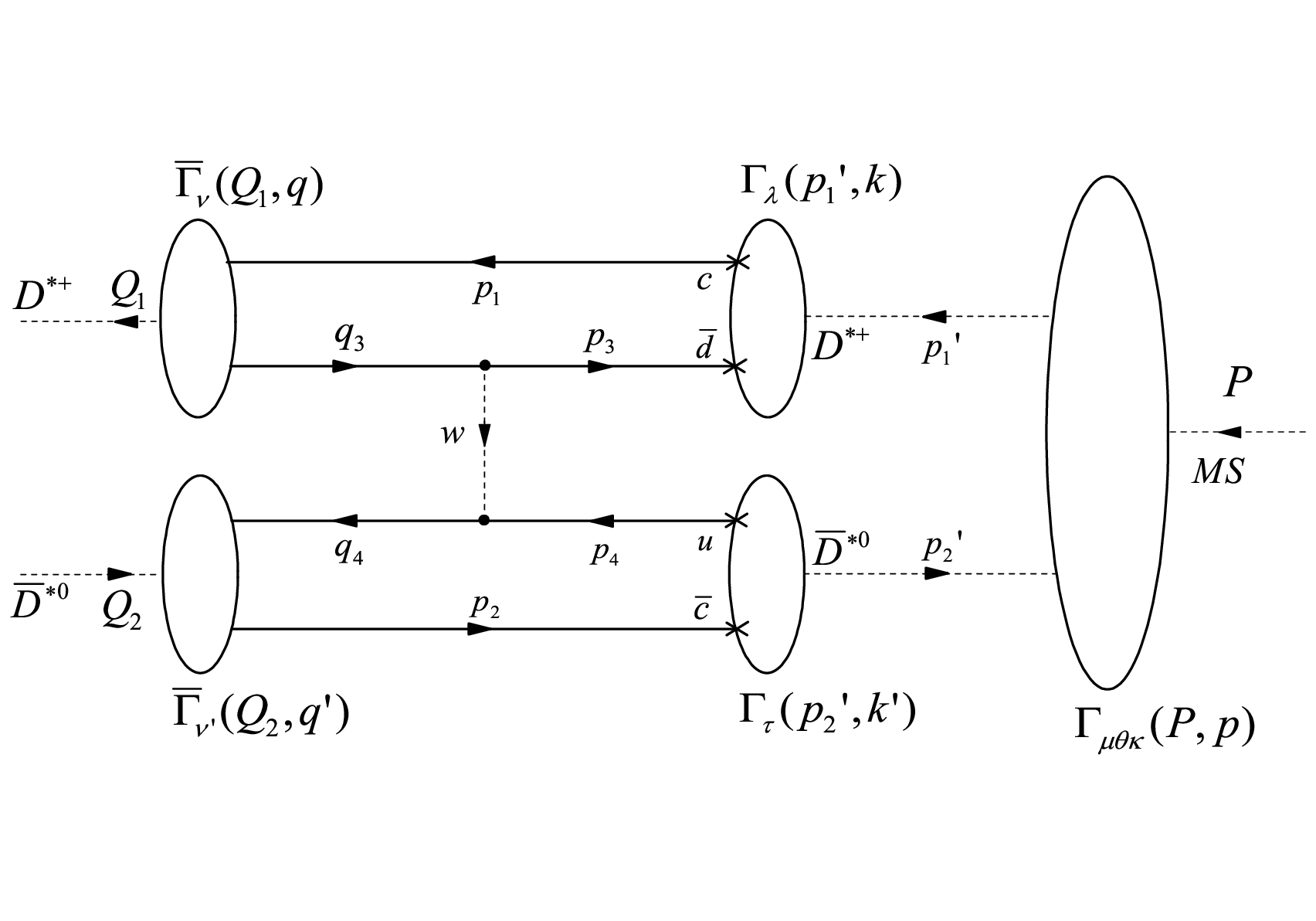}
\caption{\label{Fig3} Matrix element for channel $D^{*+}\bar D^{*0}$. The momenta in the final state satisfy $Q_1+Q_2=P^{\epsilon'}$ and the momentum of the initial state is $P$. $w$ represents the momentum of the exchanged light meson. The crosses mean that the momenta of quark propagators and the momentum $w$ of the exchanged light meson depend on the final state energy $\epsilon'$.}
\end{figure}

Considering the lowest order term of the two-particle irreducible Green's function, we obtain the interaction between two heavy vector mesons derived from one light meson exchange. Applying the generalized  Mandelstam's approach, we can obtain the $T$-matrix element $T_{(c'_2;b)a}(\epsilon')$ with arbitrary value of the final state energy $\epsilon'$ for channel $c'_2$, which can be represented graphically by Fig. \ref{Fig3}. The total energy $\epsilon'$ of the final state extends from $\epsilon_{c'_2,M}$ to $+\infty$, i.e., $\epsilon_{c'_2,M}<\epsilon'<\infty$. In Fig. \ref{Fig3}, $Q_1$ and $Q_2$ represent the momenta of final particles, $Q_1^2=-M_{D^{*+}}^2$ and $Q_2^2=-M_{\bar D^{*0}}^2$; the crosses mean that the momenta of quark propagators and the momentum $w$ of the exchanged light meson depend on $Q_1$ and $Q_2$, i.e., $p_1-p_2-p_3+p_4=p_1-p_2-q_3+q_4=Q_1+Q_2=P^{\epsilon'}$, and $p_1'-p_2'=P$. In the rest frame, we have $P=(0,0,0,iM_0)$, $P^{\epsilon'}=(0,0,0,i\epsilon')$, $\epsilon_{c'_2,M}<\epsilon'<\infty$ and $\epsilon_{c'_2,M}=M_{D^{*+}}+M_{\bar D^{*0}}$.

\begin{figure}[!htb] \centering
\includegraphics[trim = 0mm 30mm 0mm 20mm,scale=1,width=8cm]{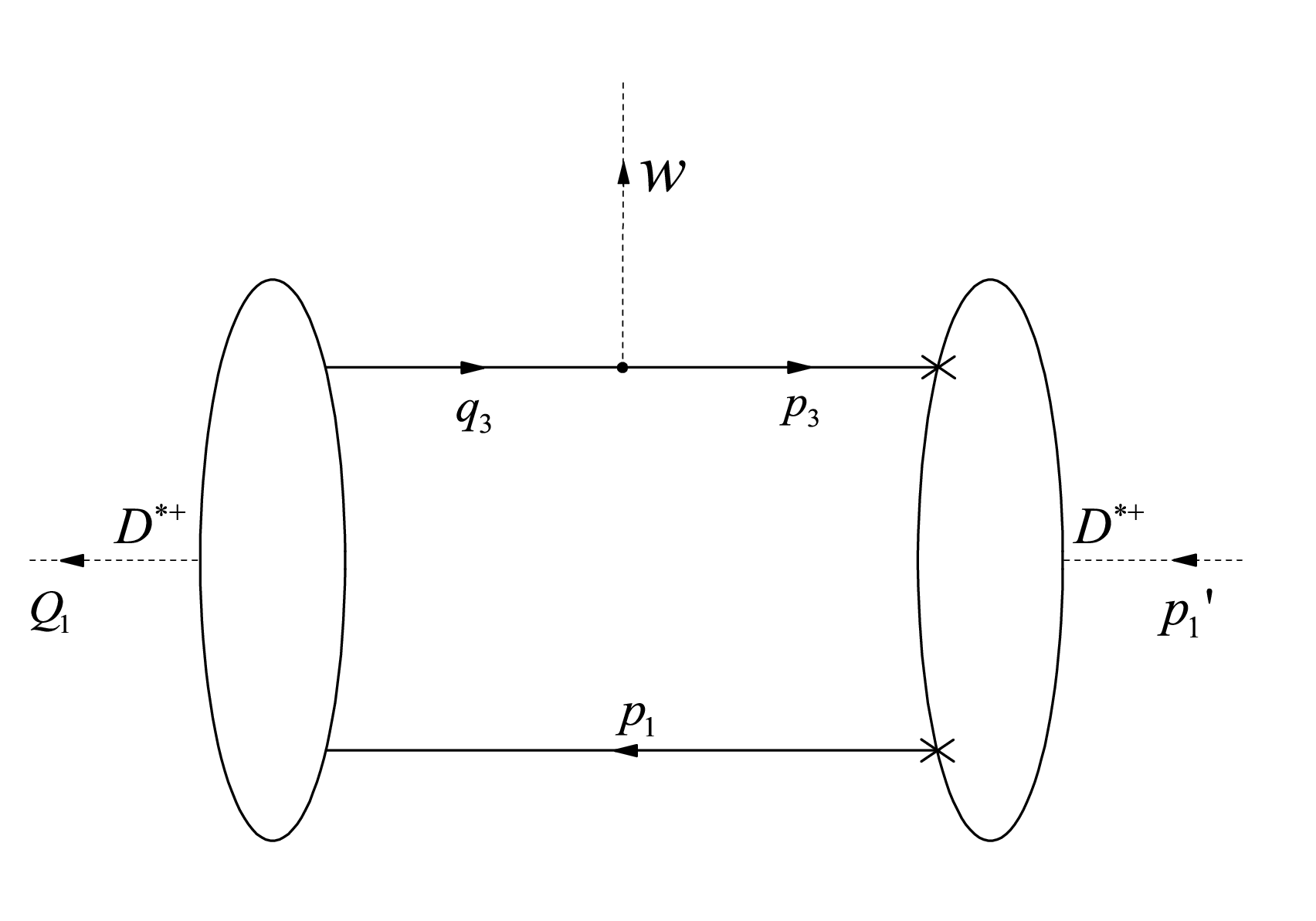}
\caption{\label{Fig4} Heavy meson form factor with respect to $\epsilon'$. $Q_1$ depends on $P^{\epsilon'}$ and $p_1'$ depends on $P$. The crosses mean that the momenta of quark propagators and the momentum $w$ of the exchanged light meson depend on the final state energy $\epsilon'$.}
\end{figure}

\begin{figure}[!htb] \centering
\includegraphics[trim = 0mm 40mm 0mm 40mm,scale=1,width=10cm]{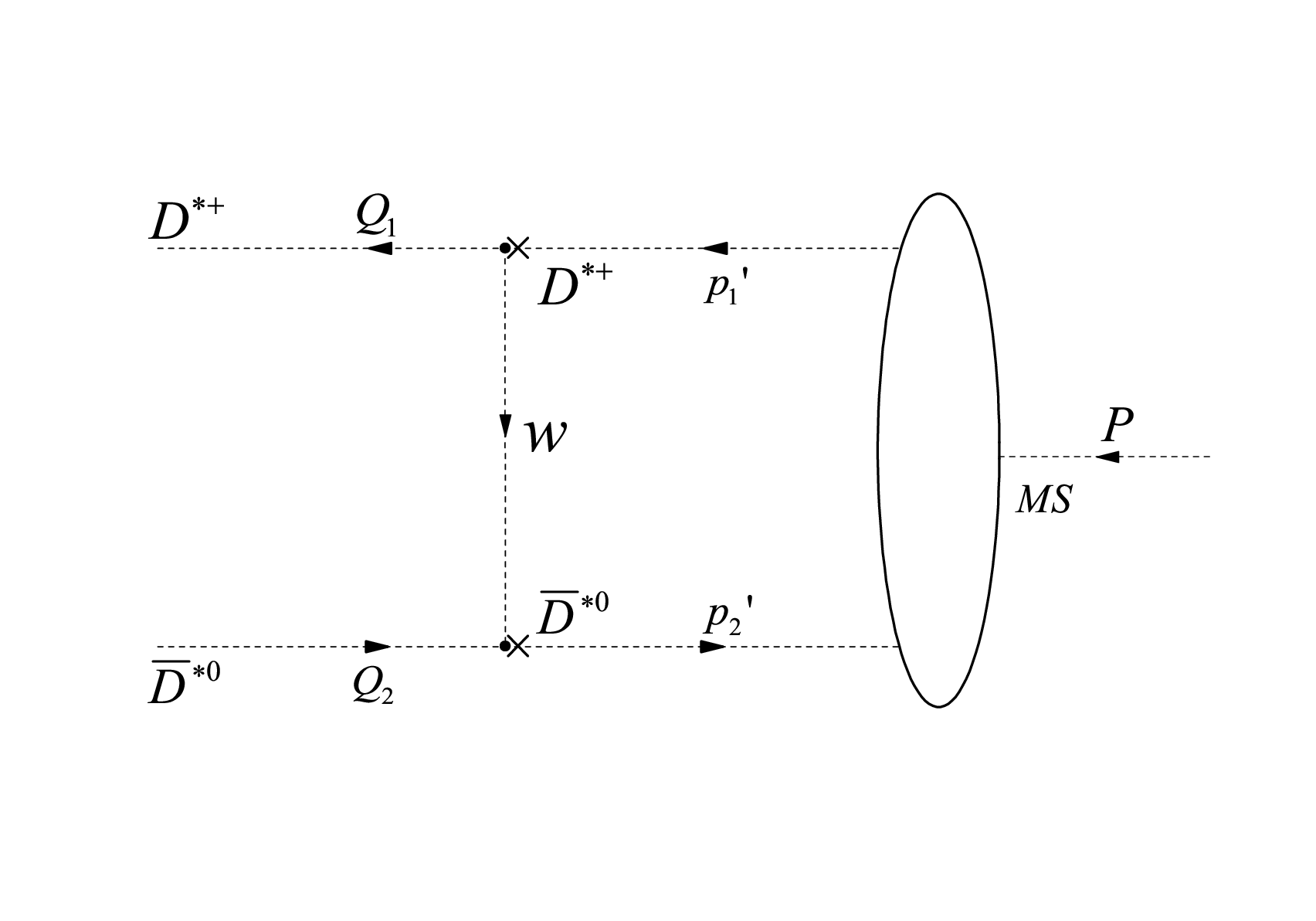}
\caption{\label{Fig5} Reduced matrix element for channel $D^{*+}\bar D^{*0}$. The crosses mean that the momentum $w$ of the exchanged light meson depends on $Q_1$ and $Q_2$.}
\end{figure}

To simplify the computational process of $T_{(c'_2;b)a}(\epsilon')$, we use the form factor of heavy meson with respect to $\epsilon'$. However, different from the ordinary form factor, we should introduce the form factor with respect to $\epsilon'$, which is shown as Fig. \ref{Fig4}. In Ref. \cite{mypaper9}, we have obtained the explicit forms for the heavy meson form factors $h(w^2)$ with respect to $\epsilon'$, and then Fig. \ref{Fig3} can be reduced to Fig. \ref{Fig5}. In Fig. \ref{Fig5}, we have $Q_1+Q_2=P^{\epsilon'}$, $p_1'-p_2'=P$, and the crosses lie on the right-hand side of light meson propagator, which implies that the momentum $w$ of the exchanged light meson depends on $Q_1$ and $Q_2$.

According to the generalized Mandelstam's approach, the $T$-matrix element becomes
\begin{equation}\label{Tmatrele2}
\begin{split}
T_{(c'_2;b)a}(\epsilon')=&\frac{-i\varepsilon_{\nu'}^{\varrho'}(Q_2)\varepsilon_\nu^{\varrho}(Q_1)\eta^\varsigma_\mu(P)}{(2\pi)^{9/2}\sqrt{2E_{\bar D^{*0}}(Q_2)}\sqrt{2E_{D^{*+}}(Q_1)}\sqrt{2E(P)}}\\
\times&\int \frac{d^4p}{(2\pi)^4}\mathcal{V}_{\nu\lambda,\tau\nu'}(Q_1,Q_2,p)\frac{1}{\mathcal{N}^{1^+}}\chi_{\mu\lambda\tau}(P,p),
\end{split}
\end{equation}
where $\varepsilon_\nu^{\varrho=1,2,3}(Q_1)$ and $\varepsilon_{\nu'}^{\varrho'=1,2,3}(Q_2)$ become the polarization vectors of $D^{*+}$ and $\bar D^{*0}$, respectively, $E_{D^*}(p)=\sqrt{\mathbf{p}^2+M^2_{D^*}}$ and $\mathcal{V}_{\nu\lambda,\tau\nu'}(Q_1,Q_2,p)$ represents the interactions derived from one light meson ($\sigma$, $\rho^0$ and $\omega$) exchange. Here $\chi_{\mu\lambda\tau}(P,p)$ is expressed as Eq. (\ref{BSWF1+}). Using the approach introduced in Ref. \cite{mypaper9}, we can obtain the interaction from one light meson ($\sigma$, $\rho^0$ and $\omega$) exchange
\begin{equation}
\begin{split}\label{decayinter}
\mathcal {V}&_{\nu\lambda,\tau\nu'}(Q_1,Q_2,p)\\
=&h_1^{(\text{s})}(w^2)\frac{-ig_\sigma^2}{w^2+M_\sigma^2}\bar{h}_1^{(\text{s})}(w^2)\delta_{\nu\lambda}\delta_{\tau\nu'}+\left(\frac{ig_\rho^2}{w^2+M_\rho^2}+\frac{-ig_\omega^2}{w^2+M_\omega^2}\right)\{h_1^{(\text{lv})}(w^2)\bar{h}_1^{(\text{lv})}(w^2)\\
&\times(Q_1+p'_1)\cdot(Q_2-p'_2)\delta_{\nu\lambda}\delta_{\tau\nu'}-h_1^{(\text{lv})}(w^2)\bar{h}_2^{(\text{lv})}(w^2)\delta_{\nu\lambda}[-(Q_1+p'_1)_\tau p'_{2\nu'}+Q_{2\tau}(Q_1+p'_1)_{\nu'}]\\
&-h_2^{(\text{lv})}(w^2)\bar{h}_1^{(\text{lv})}(w^2)[p'_{1\nu}(Q_2-p'_2)_\lambda+(Q_2-p'_2)_\nu
Q_{1\lambda}]\delta_{\tau\nu'}+h_2^{(\text{lv})}(k^2)\bar{h}_2^{(\text{lv})}(k^2)[-p'_{1\nu}\delta_{\lambda\tau}p'_{2\nu'}\\
&+p'_{1\nu}\delta_{\lambda\nu'}Q_{2\tau}-\delta_{\nu\tau}Q_{1\lambda}p'_{2\nu'}+\delta_{\nu\nu'}Q_{1\lambda}Q_{2\tau}]\},
\end{split}
\end{equation}
where $w=p-(Q_1-Q_2)/2$ is the momentum of light meson, $h(w^2)$ and $\bar h(w^2)$ are the heavy meson form factors with respect to $\epsilon'$. These momenta in Fig. \ref{Fig5} become
\begin{equation}\label{momenta3}
\begin{split}
w=(\mathbf{p}-\mathbf{Q}_{D^*}(\epsilon'),0),~~Q_1+Q_2=P^{\epsilon'},~~p_1'=p+P/2,~~p_2'=p-P/2,
\end{split}
\end{equation}
where $Q_1=(\mathbf{Q}_{D^*}(\epsilon'),i\epsilon'/2)$, $Q_2=(-\mathbf{Q}_{D^*}(\epsilon'),i\epsilon'/2)$ and $\mathbf{Q}_{D^*}^2(\epsilon')=[\epsilon'^2-(M_{D^{*+}}+M_{\bar D^{*0}})^2]/4$. From Eq. (\ref{Iepsilon'}), we obtain the function $\mathbb{I}_2(\epsilon')$ for channel $D^{*+}\bar D^{*0}$
\begin{equation}\label{Iepsilon'2}
\begin{split}
\mathbb{I}_2(\epsilon')&=\frac{1}{2}\int d^3Q_1d^3Q_2(2\pi)^4\delta^{(4)}(Q_1+Q_2-P^{\epsilon'})\frac{1}{3}\sum_{\varsigma=1}^3\sum_{\varrho'=1}^3\sum_{\varrho=1}^3|T_{(c'_2;b)a}(\epsilon')|^2.
\end{split}
\end{equation}

\section{physical mass and width of resonance}\label{sec:mw}
For resonance $T_{c\bar c}(4020)^+$, the dispersion relation (\ref{disrel}) becomes
\begin{equation}
\begin{split}
\mathbb{D}(M_0)=&-\frac{\mathcal{P}}{\pi}\int_{\epsilon_{c'_1,M}}^\infty \frac{\mathbb{I}_1(\epsilon')}{\epsilon'-M_0}d\epsilon'-\frac{1}{\pi}\int_{\epsilon_{c'_2,M}}^\infty \frac{\mathbb{I}_2(\epsilon')}{\epsilon'-M_0}d\epsilon',
\end{split}
\end{equation}
where $\epsilon_{c'_1,M}=M_{h_c}+M_\pi$, $\epsilon_{c'_2,M}=M_{D^{*+}}+M_{\bar D^{*0}}$. From Eq. (\ref{pole}), we obtain that the physical mass of resonance $T_{c\bar c}(4020)^+$ is $M=M_0+(2\pi)^3\mathbb{D}(M_0)$. Replacing $M_0$ by $M$ in Eq. (\ref{Tmatrele1}) and setting $\epsilon'=M$, we calculate the matrix element $T_{(c'_1;b)a}(M)$ and obtain that the width for physical decay model $T_{c\bar c}(4020)^+\rightarrow h_c(1P)\pi^+$ is $\Gamma_1=2(2\pi)^3\mathbb{I}_1(M)$. Since resonance $T_{c\bar c}(4020)^+$ slightly lies above the $D^{*+}\bar D^{*0}$ threshold, we should consider energy distribution to calculate the width for physical decay model $T_{c\bar c}(4020)^+\rightarrow D^{*+}\bar D^{*0}$. Replacing $M_0$ by $M$ in Eq. (\ref{Tmatrele2}) and setting $\epsilon'=M$, we calculate the matrix element $T_{(c'_2;b)a}(M)$ and obtain that the width for physical decay model $T_{c\bar c}(4020)^+\rightarrow D^{*+}\bar D^{*0}$ is $\Gamma_2=2(2\pi)^3\mathbb{I}_2(M)$.

\section{numerical result}\label{sec:nr}
Considering the isospin conservation, we employ the constituent quark masses $m_u=m_d=0.33$ GeV, the heavy quark mass $m_{c}=1.55$ GeV \cite{ts:Ebert} and the meson masses $M_{\pi^0}=M_{\pi^{\pm}}=0.135$ GeV, $M_\sigma=0.46$ GeV, $M_{\omega}=0.782$ GeV, $M_{\rho^0}=M_{\rho^{\pm}}=0.775$ GeV, $M_{D^{*0}}=M_{D^{*+}}=2.0085$ GeV, $M_{h_c}=3.525$ GeV \cite{PDG2024}. Without an adjustable parameter, we numerically solve BS equation (\ref{BSE}) and obtain the mass $M_0$ and BS wave function of bound state $D^{*+}\bar{D}^{*0}$ with spin-parity quantum numbers $1^+$. Then $M_0$ and BS wave function $\chi_{\mu\lambda\tau}(P,p)$ given in Eq. (\ref{BSWF1+}) are used to evaluate the $T$-matrix elements with respect to arbitrary value of the final state energy $\epsilon'$. From Eqs. (\ref{Tmatrele1}) and (\ref{momenta}), we calculate the $T$-matrix element $T_{(c'_1;b)a}(\epsilon')$ with respect to $\epsilon'$ for channel $h_c(1P)\pi^+$. From Eqs. (\ref{Tmatrele2}), (\ref{decayinter}) and (\ref{momenta3}), we calculate the $T$-matrix element $T_{(c'_2;b)a}(\epsilon')$ with respect to $\epsilon'$ for channel $D^{*+}\bar D^{*0}$. From Eqs. (\ref{Iepsilon'1}) and (\ref{Iepsilon'2}), we calculate the functions $\mathbb{I}_1(\epsilon')$ over $\epsilon_{c'_1,M}<\epsilon'<\infty$ and $\mathbb{I}_2(\epsilon')$ over $\epsilon_{c'_2,M}<\epsilon'<\infty$, respectively. By doing the numerical calculation, we obtain the mass correction $\Delta M=(2\pi)^3\mathbb{D}(M_0)$ and the physical mass $M$ for resonance $T_{c\bar c}(4020)^+$. Finally, the physical mass is used to recalculate these strong decay widths $\Gamma_1(T_{c\bar c}(4020)^+\rightarrow h_c(1P)\pi^+)$ and $\Gamma_2(T_{c\bar c}(4020)^+\rightarrow D^{*+}\bar D^{*0})$. $M$ and $\Gamma$ should be the observed mass and full width in experiments and our numerical results for resonance $T_{c\bar c}(4020)^+$ are in good agreement with the experimental data, which are presented in Table \ref{table1}.
\newcommand{\tabincell}[2]{\begin{tabular}{@{}#1@{}}#2\end{tabular}}
\begin{table}[ht]
\caption{Mass $M$ and full width $\Gamma$ for physical resonance $T_{c\bar c}(4020)^+$. $M_0$ is the mass of bound state $D^{*+}\bar{D}^{*0}$, $\Delta M$ is the calculated correction due to all channels, and $\Gamma_i$ is the calculated width of $i$th decay channel. (Dimensioned quantities in MeV.)}
\label{table1} \vspace*{-6pt}
\begin{center}
\begin{tabular}{ccccccc}\hline
Quantity  & ~~$M_0$ & ~~$\Delta M$ & ~~$M$ & ~~~$\Gamma_1$ & ~~~$\Gamma_2$  & ~~$\Gamma$
\\ \hline
This work     & ~~4016.15 & ~~2.86 &  ~~4019.01  & ~~~14.32 & ~~~$0.2\times10^{-3}$ &~~14.32
\\
\hline
PDG\cite{PDG2024} & & & ~~4024.1$\pm$1.9 & & & ~~13$\pm$5
\\
\hline
\end{tabular}
\end{center}
\end{table}

In Table \ref{table1}, the calculated mass $M_0$ ($M_0<M_{D^{*+}}+M_{\bar D^{*0}}$) of meson-meson bound state $D^{*+}\bar D^{*0}$ should not be the mass of physical resonance $T_{c\bar c}(4020)^+$. After considering the correction for energy level of molecular state due to decay channels, we obtain the physical mass $M$ ($M>M_{D^{*+}}+M_{\bar D^{*0}}$) of resonance $T_{c\bar c}(4020)^+$, which is consistent with experimental value. Therefore, it is more reasonable to regard exotic resonance $T_{c\bar c}(4020)^+$ as an unstable meson-meson molecular state $D^{*+}\bar{D}^{*0}$. This work provides a further verification for the molecular hypothesis of $T_{c\bar c}(4020)^+$ with spin-parity quantum numbers $1^+$ and we can draw the conclusion that the decay width $\Gamma_2(T_{c\bar c}(4020)^+\rightarrow D^{*+}\bar D^{*0})$ is very small compared with the decay width $\Gamma_1(T_{c\bar c}(4020)^+\rightarrow h_c(1P)\pi^+)$.

In the actual calculation, we require the meson-quark coupling constants $g$ and the parameters in BS amplitudes of heavy mesons to calculate the mass and decay width of physical resonance. The meson-quark coupling constants can be determined by QCD sum rules approach \cite{cc2}, the parameter in BS amplitude of heavy meson $D^*$ is fixed by providing fits to observables \cite{BSE:Roberts4,BSE:Roberts5}, and the parameters in BS wave function of heavy axial-vector meson $h_c(1P)$ are determined by fitting homogeneous Bethe-Salpeter equation of $h_c(1P)$. Our approach also involves the constituent quark masses $m_u$, $m_d$, and the heavy quark mass $m_c$. According to the spontaneous breaking of chiral symmetry, the light quarks ($u,d,s$) obtain their constituent masses because the vacuum condensate is not equal to zero, and the heavy quark mass $m_c$ is irrelevant to vacuum condensate. Normally, the value slightly greater than a third of nucleon mass is employed as the constituent mass of light quark. The value of heavy quark mass $m_c$ can be determined by the experimental mass of charmonium system $J/\psi$. The constituent masses of light quarks have not been exactly determined, while the large uncertainty should not exist for heavy quark. Taking different values for constituent masses of light quarks $m_{u,d}$, we do the numerical calculation again and obtain the uncertainties given in Table \ref{table2}. These numerical results imply that in our approach the calculated masses and decay widths depend on the values of constituent quark masses, but not sensitively. Despite the large uncertainty of meson mass $M_\sigma$, it has been found that the uncertainties of numerical results from meson mass $M_\sigma$ are also very small in our previous works \cite{mypaper4,mypaper5,mypaper6,mypaper7} and Refs. \cite{Msigma1,Msigma2}.
\begin{table}[ht]
\caption{Mass $M$ and full width $\Gamma$ for physical resonance $T_{c\bar c}(4020)^+$. We vary constituent quark masses to estimate the uncertainties. (Dimensioned quantities in MeV.)}
\label{table2} \vspace*{-6pt}
\begin{center}
\begin{tabular}{ccccccc}\hline
Quantity  & ~~$M_0$ & ~~$\Delta M$ & ~~$M$ & ~~~$\Gamma_1$ & ~~~$\Gamma_2$  & ~~$\Gamma$
\\ \hline
\tabincell{c}{$m_{u,d}=320$}     & ~~4007.44 & ~~2.29 &  ~~4009.73  & ~~~11.38 & ~~~$0.2\times10^{-3}$ &~~11.38
\\ \hline
\tabincell{c}{$m_{u,d}=330$}     & ~~4016.15 & ~~2.86 &  ~~4019.01  & ~~~14.32 & ~~~$0.2\times10^{-3}$ &~~14.32
\\
\hline
\tabincell{c}{$m_{u,d}=340$}     & ~~4024.78 & ~~3.31 &  ~~4028.09  & ~~~18.09 & ~~~$0.2\times10^{-3}$ &~~18.09
\\
\hline
\end{tabular}
\end{center}
\end{table}

\section{conclusion}\label{sec:concl}
Since the observed resonance $T_{c\bar c}(4020)$ lies above the $D^{*}\bar{D}^{*}$ threshold, in principle one can not explain $T_{c\bar c}(4020)$ as a meson-meson bound state. In this work, exotic resonance $T_{c\bar c}(4020)$ is considered as an unstable molecular state $D^{*}\bar{D}^{*}$ with spin-parity quantum numbers $1^+$, and we investigate the time evolution of meson-meson molecular state as determined by the total Hamiltonian. According to the developed Bethe-Salpeter theory, we calculate the correction for energy level of molecular state due to decay channels and find that the unstable meson-meson molecular state $D^{*}\bar D^{*}$ lies above the threshold. Then the physical mass of resonance $T_{c\bar c}(4020)$ is used to calculate its decay widths, which are in good agreement with the experimental data. Therefore, it is more reasonable to regard exotic resonance $T_{c\bar c}(4020)$ as an unstable meson-meson molecular state $D^{*}\bar{D}^{*}$.

\begin{acknowledgements}
This work was supported by the National Natural Science Foundation of China under Grants No. 11705104, No. 11801323 and No. 52174145; Shandong Provincial Natural Science Foundation, China under Grants No. ZR2023MA083, No. ZR2016AQ19 and No. ZR2016AM31; and SDUST Research Fund under Grant No. 2018TDJH101.
\end{acknowledgements}

\appendix

\section{Tensor structures in the general form of BS wave functions}\label{app1}
The tensor structures in Eqs. (\ref{jp}) and (\ref{jm}) are given below \cite{mypaper9}
\begin{equation*}
\mathcal{T}_{\lambda\tau}^1=(p^2+\eta_1P\cdot p-\eta_2P\cdot p-\eta_1\eta_2P^2)g_{\lambda\tau}-(p_{\lambda}p_{\tau}+\eta_1P_{\tau}p_{\lambda}-\eta_2P_{\lambda}p_{\tau}-\eta_1\eta_2P_{\lambda}P_{\tau}),
\end{equation*}
\begin{equation*}
\begin{split}
\mathcal{T}_{\lambda\tau}^2=&(p^2+2\eta_1P\cdot p+\eta_1^2P^2)(p^2-2\eta_2P\cdot p+\eta_2^2P^2)g_{\lambda\tau}\\
&+(p^2+\eta_1P\cdot p-\eta_2P\cdot p-\eta_1\eta_2P^2)(p_{\lambda}p_{\tau}+\eta_1P_{\lambda}p_{\tau}-\eta_2P_{\tau}p_{\lambda}-\eta_1\eta_2P_{\lambda}P_{\tau})\\
&-(p^2-2\eta_2P\cdot p+\eta_2^2P^2)(p_{\lambda}p_{\tau}+\eta_1P_{\lambda}p_{\tau}+\eta_1P_{\tau}p_{\lambda}+\eta_1 ^2P_{\lambda}P_{\tau})\\
&-(p^2+2\eta_1P\cdot p+\eta_1^2P^2)(p_{\lambda}p_{\tau}-\eta_2P_{\lambda}p_{\tau}-\eta_2P_{\tau}p_{\lambda}+\eta_2 ^2P_{\lambda}P_{\tau}),
\end{split}
\end{equation*}
\begin{equation*}
\begin{split}
\mathcal{T}_{\mu_1\cdots\mu_j\lambda\tau}^3=&\frac{1}{j!}p_{\{\mu_2}\cdots p_{\mu_j}g_{\mu_1\}\lambda}(p^2+2\eta_1P\cdot p+\eta_1^2P^2)[(p^2-2\eta_2P\cdot p+\eta_2^2P^2)(p+\eta_1P)_{\tau}\\
&-(p^2+\eta_1P\cdot p-\eta_2P\cdot p-\eta_1\eta_2P^2)(p-\eta_2P)_{\tau}]\\
&-p_{\mu_1}\cdots p_{\mu_j}[(p^2-2\eta_2P\cdot p+\eta_2^2P^2)(p_{\lambda}p_{\tau}+\eta_1P_{\lambda}p_{\tau}+\eta_1P_{\tau}p_{\lambda}+\eta_1 ^2P_{\lambda}P_{\tau})\\
&-(p^2+\eta_1P\cdot p-\eta_2P\cdot p-\eta_1\eta_2P^2)(p_{\lambda}p_{\tau}+\eta_1P_{\lambda}p_{\tau}-\eta_2P_{\tau}p_{\lambda}-\eta_1\eta_2P_{\lambda}P_{\tau})],
\end{split}
\end{equation*}
\begin{equation*}
\begin{split}
\mathcal{T}_{\mu_1\cdots\mu_j\lambda\tau}^4=&\frac{1}{j!}p_{\{\mu_2}\cdots p_{\mu_j}g_{\mu_1\}\tau}(p^2-2\eta_2P\cdot p+\eta_2^2P^2)[(p^2+\eta_1P\cdot p\\
&-\eta_2P\cdot p-\eta_1\eta_2P^2)(p+\eta_1P)_{\lambda}-(p^2+2\eta_1P\cdot p+\eta_1^2P^2)(p-\eta_2P)_{\lambda}]\\
&+p_{\mu_1}\cdots p_{\mu_j}[(p^2+2\eta_1P\cdot p+\eta_1^2P^2)(p_{\lambda}p_{\tau}-\eta_2P_{\lambda}p_{\tau}-\eta_2P_{\tau}p_{\lambda}+\eta_2 ^2P_{\lambda}P_{\tau})\\
&-(p^2+\eta_1P\cdot p-\eta_2P\cdot p-\eta_1\eta_2P^2)(p_{\lambda}p_{\tau}+\eta_1P_{\lambda}p_{\tau}-\eta_2P_{\tau}p_{\lambda}-\eta_1\eta_2P_{\lambda}P_{\tau})],
\end{split}
\end{equation*}
\begin{equation*}
\begin{split}
\mathcal{T}^5_{\mu_1\cdots\mu_j\lambda\tau}=&\frac{1}{j!}(p^2+2\eta_1P\cdot p+\eta_1^2P^2)(p^2-2\eta_2P\cdot p+\eta_2^2P^2)p_{\{\mu_3}\cdots p_{\mu_j}g_{\mu_1\lambda}g_{\mu_2\}\tau}\\
&-\frac{1}{j!}p_{\{\mu_2}\cdots p_{\mu_j}g_{\mu_1\}\tau}(p^2-2\eta_2P\cdot p+\eta_2^2P^2)(p+\eta_1P)_{\lambda}\\
&-\frac{1}{j!}p_{\{\mu_2}\cdots p_{\mu_j}g_{\mu_1\}\lambda}(p^2+2\eta_1P\cdot p+\eta_1^2P^2)(p-\eta_2P)_{\tau}\\
&+p_{\mu_1}\cdots p_{\mu_j}(p_{\lambda}p_{\tau}+\eta_1P_{\lambda}p_{\tau}-\eta_2P_{\tau}p_{\lambda}-\eta_1\eta_2P_{\lambda}P_{\tau}),
\end{split}
\end{equation*}
\begin{equation*}
\begin{split}
\mathcal{T}^6_{\mu_1\cdots\mu_j\lambda\tau}=&p_{\{\mu_3}\cdots p_{\mu_j}\epsilon_{\mu_{1}\lambda\xi\zeta}p_\xi P_\zeta\epsilon_{\mu_{2}\}\tau\xi'\zeta'}p_{\xi'}P_{\zeta'},
\end{split}
\end{equation*}
\begin{equation*}
\begin{split}
\mathcal{T}_{\mu_1\cdots\mu_j\lambda\tau}^7=&-(2p^2+\eta_1P\cdot p-\eta_2P\cdot p)p_{\{\mu_2}\cdots p_{\mu_j}\epsilon_{\mu_1\}\lambda\tau\xi}p_\xi\\
&+(2\eta_1\eta_2P\cdot p+\eta_2p^2-\eta_1p^2)p_{\{\mu_2}\cdots p_{\mu_j}\epsilon_{\mu_1\}\lambda\tau\xi}P_\xi\\
&+p_{\{\mu_2}\cdots p_{\mu_j}\epsilon_{\mu_1\}\lambda\xi\zeta}p_\xi P_\zeta p_\tau+p_{\{\mu_2}\cdots p_{\mu_j}\epsilon_{\mu_1\}\tau\xi\zeta}p_\xi P_\zeta p_\lambda,
\end{split}
\end{equation*}
\begin{equation*}\
\begin{split}
\mathcal{T}_{\mu_1\cdots\mu_j\lambda\tau}^8=&-(P\cdot p)p_{\{\mu_2}\cdots p_{\mu_j}\epsilon_{\mu_1\}\lambda\tau\xi}p_\xi+p^2p_{\{\mu_2}\cdots p_{\mu_j}\epsilon_{\mu_1\}\lambda\tau\xi}P_\xi\\
&-p_{\{\mu_2}\cdots p_{\mu_j}\epsilon_{\mu_1\}\lambda\xi\zeta}p_\xi P_\zeta p_\tau+p_{\{\mu_2}\cdots p_{\mu_j}\epsilon_{\mu_1\}\tau\xi\zeta}p_\xi P_\zeta p_\lambda,
\end{split}
\end{equation*}
\begin{equation*}
\begin{split}
\mathcal{T}_{\mu_1\cdots\mu_j\lambda\tau}^{9}=&-(2P\cdot p+\eta_1P^2-\eta_2P^2)p_{\{\mu_2}\cdots p_{\mu_j}\epsilon_{\mu_1\}\lambda\tau\xi}p_\xi\\
&+P\cdot(\eta_2p-\eta_1p+2\eta_1\eta_2P)p_{\{\mu_2}\cdots p_{\mu_j}\epsilon_{\mu_1\}\lambda\tau\xi}P_\xi\\
&+p_{\{\mu_2}\cdots p_{\mu_j}\epsilon_{\mu_1\}\lambda\xi\zeta}p_\xi P_\zeta P_\tau+p_{\{\mu_2}\cdots p_{\mu_j}\epsilon_{\mu_1\}\tau\xi\zeta}p_\xi P_\zeta P_\lambda,
\end{split}
\end{equation*}
\begin{equation*}
\begin{split}
\mathcal{T}_{\mu_1\cdots\mu_j\lambda\tau}^{10}=&-P^2p_{\{\mu_2}\cdots p_{\mu_j}\epsilon_{\mu_1\}\lambda\tau\xi}p_\xi+(P\cdot p)p_{\{\mu_2}\cdots p_{\mu_j}\epsilon_{\mu_1\}\lambda\tau\xi}P_\xi\\
&-p_{\{\mu_2}\cdots p_{\mu_j}\epsilon_{\mu_1\}\lambda\xi\zeta}p_\xi P_\zeta P_\tau+p_{\{\mu_2}\cdots p_{\mu_j}\epsilon_{\mu_1\}\tau\xi\zeta}p_\xi P_\zeta P_\lambda,
\end{split}
\end{equation*}
\begin{equation*}
\begin{split}
\mathcal{T}_{\mu_1\cdots\mu_j\lambda\tau}^{11}=&(p^2+\eta_1P\cdot p-\eta_2P\cdot p-\eta_1\eta_2P^2)p_{\{\mu_3}\cdots p_{\mu_j}g_{\mu_1\lambda}\epsilon_{\mu_2\}\tau\xi\zeta}p_\xi P_\zeta\\
&-p_{\{\mu_2}\cdots p_{\mu_j}\epsilon_{\mu_1\}\tau\xi\zeta}p_\xi P_\zeta(p-\eta_2P)_\lambda,
\end{split}
\end{equation*}
\begin{equation*}
\begin{split}
\mathcal{T}_{\mu_1\cdots\mu_j\lambda\tau}^{12}=&(p^2+\eta_1P\cdot p-\eta_2P\cdot p-\eta_1\eta_2P^2)p_{\{\mu_3}\cdots p_{\mu_j}g_{\mu_1\tau}\epsilon_{\mu_2\}\lambda\xi\zeta}p_\xi P_\zeta\\
&-p_{\{\mu_2}\cdots p_{\mu_j}\epsilon_{\mu_1\}\lambda\xi\zeta}p_\xi P_\zeta(p+\eta_1P)_\tau,
\end{split}
\end{equation*}
where $\{\mu_1,\cdots,\mu_j\}$ represents symmetrization of the indices $\mu_1,\cdots,\mu_j$.

\section{Form factors in instantaneous approximation}\label{app2}
From the effective interaction Lagrangian (\ref{Lag}), one can obtain the quark current $J_\alpha$ coupling with light vector meson, the quark pseudoscalar density $J^-$ coupling with light pseudoscalar meson and the quark scalar density $J$ coupling with $\sigma$ meson. From the Lorentz-structure, the matrix elements of quark pseudoscalar density $J^-$, quark scalar density $J$ and quark current $J_\alpha$ can be expressed as
\begin{subequations}\label{vertice}
\begin{eqnarray}
\begin{split}
\langle VM^\varrho(p_1')|J^-(0)|VM^\vartheta(q_1') \rangle=\frac{1}{2\sqrt{E_{D^{*+}}(p_1')E_{D^{*+}}(q_1')}}h^{(\text{p})}(w^{2})\epsilon_{\varsigma\varsigma'\theta\theta'}p_{1\varsigma}'q_{1\varsigma'}'\varepsilon^\varrho_\theta(p_1')\varepsilon^\vartheta_{\theta'}(q_1'),
\end{split}
\end{eqnarray}
\begin{eqnarray}
\begin{split}
\langle \overline{VM'}^{\varrho'}(-p_2')|J^-(0)|\overline{VM'}^{\vartheta'}(-q_2') \rangle=&\frac{1}{2\sqrt{E_{\bar D^{*0}}(-p_2')E_{\bar D^{*0}}(-q_2')}}\\
&\times\bar h^{(\text{p})}(w^{2})\epsilon_{\varpi\varpi'\kappa'\kappa}p_{2\varpi}'q_{2\varpi'}'\varepsilon^{\varrho'}_\kappa(-p_2')\varepsilon^{\vartheta'}_{\kappa'}(-q_2'),
\end{split}
\end{eqnarray}
\begin{equation}
\begin{split}
\langle &VM^\varrho(p_1')|J(0)|VM^\vartheta(q_1') \rangle=\frac{1}{2\sqrt{E_{D^{*+}}(p_1')E_{D^{*+}}(q_1')}}\\
&\times\bigg\{[\varepsilon^\varrho(p_1')\cdot\varepsilon^\vartheta(q_1')]h^{(\text{s})}_{1}(w^{2})-h^{(\text{s})}_{2}(w^{2})\frac{1}{M_{1}^{2}}[\varepsilon^\varrho(p_1')\cdot q_1'][\varepsilon^\vartheta(q_1')\cdot p_1']\bigg\},
\end{split}
\end{equation}
\begin{equation}
\begin{split}
\langle &\overline{VM'}^{\varrho'}(-p_2')|J(0)|\overline{VM'}^{\vartheta'}(-q_2')\rangle=\frac{1}{2\sqrt{E_{\bar D^{*0}}(-p_2')E_{\bar D^{*0}}(-q_2')}}\\
&\times\bigg\{[\varepsilon^{\varrho'}(-p_2')\cdot\varepsilon^{\vartheta'}(-q_2')]\bar{h}^{(\text{s})}_{1}(w^{2})-\bar{h}^{(\text{s})}_{2}(w^{2})\frac{1}{M_{2}^{2}}[\varepsilon^{\varrho'}(-p_2')\cdot(-q_2')][\varepsilon^{\vartheta'}(-q_2')\cdot (-p_2')]\bigg\},
\end{split}
\end{equation}
\begin{equation}\label{verticec}
\begin{split}
\langle &VM^\varrho(p_1')|J_{\alpha}(0)|VM^\vartheta(q_1')\rangle=\frac{1}{2\sqrt{E_{D^{*+}}(p_1')E_{D^{*+}}(q_1')}}\\
&\times\bigg\{[\varepsilon^\varrho(p_1')\cdot\varepsilon^\vartheta(q_1')]h_{1}^{(\text{lv})}(w^{2})(p_1'+q_1')_{\alpha}-h^{(\text{lv})}_{2}(w^{2})\{[\varepsilon^\varrho(p_1')\cdot q_1']\varepsilon^\vartheta_{\alpha}(q_1')\\
&+[\varepsilon^\vartheta(q_1')\cdot p_1']\varepsilon^\varrho_{\alpha}(p_1')\}-h_{3}^{(\text{lv})}(w^{2})\frac{1}{M_1^{2}}[\varepsilon^\varrho(p_1')\cdot q_1'][\varepsilon^\vartheta(q_1')\cdot p_1'](p_1'+q_1')_{\alpha}\bigg\},
\end{split}
\end{equation}
\begin{equation}\label{verticed}
\begin{split}
\langle &\overline{VM'}^{\varrho'}(-p_2')|J_{\beta}(0)|\overline{VM'}^{\vartheta'}(-q_2')\rangle=\frac{1}{2\sqrt{E_{\bar D^{*0}}(-p_2')E_{\bar D^{*0}}(-q_2')}}\\
&\times\bigg\{[\varepsilon^{\varrho'}(-p_2')\cdot\varepsilon^{\vartheta'}(-q_2')]\bar{h}_{1}^{(\text{lv})}(w^{2})(-p_2'-q_2')_{\beta}\\
&-\bar{h}^{(\text{lv})}_{2}(w^{2})\{[\varepsilon^{\varrho'}(-p_2')\cdot (-q_2')]\varepsilon^{\vartheta'}_{\beta}(-q_2')+[\varepsilon^{\vartheta'}(-q_2')\cdot (-p_2')]\varepsilon^{\varrho'}_{\beta}(-p_2')\}\\
&-\bar{h}_{3}^{(\text{lv})}(w^{2})\frac{1}{M_2^{2}}[\varepsilon^{\varrho'}(-p_2')\cdot (-q_2')][\varepsilon^{\vartheta'}(-q_2')\cdot (-p_2')](-p_2'-q_2')_{\beta}\bigg\},
\end{split}
\end{equation}
\end{subequations}
where $VM$ represents the vector meson $D^{*+}$, $\overline{VM'}$ represents the anti-particle of vector meson $D^{*0}$, $p'_1=(\mathbf{p},ip_{10}')$, $p'_2=(\mathbf{p},ip_{20}')$, $q'_1=(\mathbf{p}',iq_{10}')$, $q'_2=(\mathbf{p}',iq_{20}')$, $w=q_1'-p_1'=q_2'-p_2'$ is the momentum of the exchanged meson and $\mathbf{w}=\mathbf{p}'-\mathbf{p}$; $h(w^2)$ and $\bar h(w^2)$ are scalar functions, the four-vector $\varepsilon(p)$ is the polarization vector of heavy vector meson with momentum $p$ and $E_{D^*}(p)=\sqrt{\mathbf{p}^{2}+M_{D^*}^{2}}$. Taking away the external lines including normalizations and polarization vectors $\varepsilon^\varrho_\theta(p_1')$, $\varepsilon^\vartheta_{\theta'}(q_1')$, $\varepsilon^{\varrho'}_\kappa(-p_2')$, $\varepsilon^{\vartheta'}_{\kappa'}(-q_2')$, we can obtain the interaction kernel from one light meson ($\pi^0$, $\eta$, $\sigma$, $\rho^0$ and $\omega$) exchange \cite{mypaper3,mypaper4}.

\begin{figure}[!htb] \centering
\includegraphics[trim = 0mm 30mm 0mm 20mm,scale=1,width=8cm]{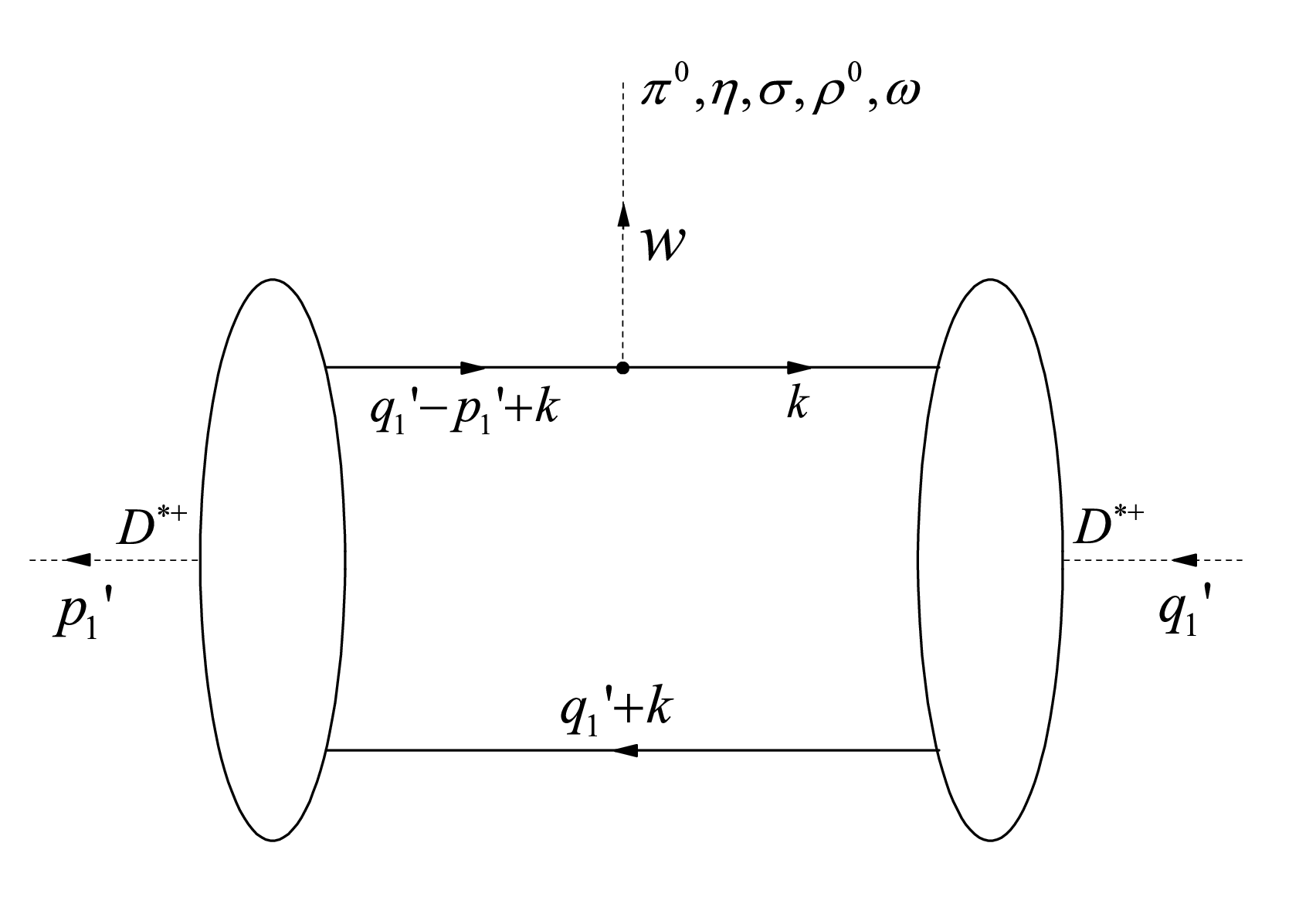}
\caption{\label{Fig6} Heavy meson form factor coupling to light meson.}
\end{figure}

In our approach, we consider the lowest order vertex to calculate these heavy vector meson form factors $h(w^2)$ describing the heavy meson structure, which is shown as Fig. \ref{Fig6}. Let $D^*_l$ denote one of $D^{*0}$ and $D^{*+}$, and $l=u,d$ represents the $u$ or $d$ antiquark in heavy vector meson $D^{*0}$ or $D^{*+}$, respectively; $\bar{D}_l^{*}$ denotes the antiparticle of $D_l^{*}$. From BS wave function of $D^*$ meson expressed as Eq. (\ref{D0BSwf}), we have obtained the heavy vector meson wave function in instantaneous approximation \cite{mypaper9}
\begin{equation}\label{wfV}
\begin{split}
\Psi_{ij}^{D_l^*}(\mathbf{k})&=\int dk_4\frac{1}{\mathcal{N}^{D_l^*}}\exp\bigg(\frac{-\mathbf{k}^2-k_4^2}{\omega_{D_l^*}^2}\bigg)\frac{\mathbf{k}^2/3+k_4^2+m_cm_l}
{(\mathbf{k}^2+k_4^2+m_c^2)(\mathbf{k}^2+k_4^2+m_l^2)}\delta_{ij}~~i,j=1,2,3.
\end{split}
\end{equation}
In instantaneous approximation, we used the method introduced in Ref. \cite{ts:Ebert} and obtained the form factors for the vertices of heavy vector meson $D_l^*$ coupling to pseudoscalar meson ($\pi$ and $\eta$)  \cite{mypaper3,mypaper4,mypaper5}
\begin{eqnarray}\label{ffp}
\begin{split}
h^{(\text{p})}(w^{2})=\bar h^{(\text{p})}(w^{2})=0,
\end{split}
\end{eqnarray}
to scalar meson ($\sigma$)
\begin{equation}\label{ffs}
\begin{split}
-\frac{h_{1}^{(\text{s})}(w^{2})}{2E_1}=&\frac{\bar{h}_{1}^{(\text{s})}(w^{2})}{2E_2}=F_1({\mathbf{w}^{2}}),~~~h_{2}^{(\text{s})}(w^{2})=\bar{h}_{2}^{(\text{s})}(w^{2})=0,\\
F_{1}(\mathbf{w}^{2})=&\int \frac{d^{3}k}{(2\pi)^{3}}\bar{\Psi}^{D_l^*}\bigg(\mathbf{k}+\frac{2E_c(k)}{E_{D_l^*}+M_{D_l^*}}\mathbf{w}\bigg)\sqrt{\frac{E_l(k)+m_l}{E_l(k+w)+m_l}}\\
&\times\left\{\frac{E_l(k+w)-E_l(k)+2m_l}{2\sqrt{E_l(k+w)E_l(k)}}-\frac{\mathbf{k\cdot w}}{2\sqrt{E_l(k+w)E_l(k)}[E_l(k)+m_l]}\right\}\Psi^{D_l^*}(\mathbf{k}),
\end{split}
\end{equation}
and to light vector meson ($\rho$ and $\omega$)
\begin{equation}\label{ffv}
\begin{split}
h_{1}^{(\text{lv})}(w^{2})=&h_{2}^{(\text{lv})}(w^{2})=\bar{h}_{1}^{(\text{lv})}(w^{2})=\bar{h}_{2}^{(\text{lv})}(w^{2})=F^{(\text{lv})}_2({\mathbf{w}^{2}}),~~~h_{3}^{(\text{lv})}(w^{2})=\bar{h}_{3}^{(\text{lv})}(w^{2})=0,\\
F^{(\text{lv})}_2(\mathbf{w}^{2})=&\frac{2\sqrt{E_{D_l^*}M_{D_l^*}}}{E_{D_l^*}+M_{D_l^*}}\int\frac{d^{3}k}{(2\pi)^{3}}\bar{\Psi}^{D_l^*}\bigg(\mathbf{k}+\frac{2E_c(k)}{E_{D_l^*}+M_{D_l^*}}\mathbf{w}\bigg)\sqrt{\frac{E_l(k)+m_l}{E_l(k+w)+m_l}}\\
&\times\left\{\frac{E_l(k+w)+E_l(k)}{2\sqrt{E_l(k+w)E_l(k)}}+\frac{\mathbf{k\cdot w}}{2\sqrt{E_l(k+w)E_l(k)}[E_l(k)+m_l]}\right\}\Psi^{D_l^*}(\mathbf{k}),
\end{split}
\end{equation}
where $E_{c,l}(p)=\sqrt{\mathbf{p}^{2}+m_{c,l}^{2}}$ and $\Psi^{D_l^*}$ is the wave function of heavy vector meson expressed as Eq. (\ref{wfV}).

Using the method above, we have obtained scalar functions for the vertex of heavy vector meson $D_l^*$ coupling to $J/\psi$ \cite{mypaper9}
\begin{equation}\label{ffhv}
\begin{split}
h_{1}^{(\text{hv})}(w^{2})=&h_{2}^{(\text{hv})}(w^{2})=\bar{h}_{1}^{(\text{hv})}(w^{2})=\bar{h}_{2}^{(\text{hv})}(w^{2})=F^{(\text{hv})}_2({\mathbf{w}^{2}}),h_{3}^{(\text{hv})}(w^{2})=\bar{h}_{3}^{(\text{hv})}(w^{2})=0,\\
F^{(\text{hv})}_2(\mathbf{w}^{2})=&\frac{2\sqrt{E_{D_l^*}M_{D_l^*}}}{E_{D_l^*}+M_{D_l^*}}\int\frac{d^{3}k}{(2\pi)^{3}}\bar{\Psi}^{D_l^*}\bigg(\mathbf{k}-\frac{2E_l(k)}{E_{D_l^*}+M_{D_l^*}}\mathbf{w}\bigg)\sqrt{\frac{E_c(k)+m_c}{E_c(k-w)+m_c}}\\
&\times\left\{\frac{E_c(k-w)+E_c(k)}{2\sqrt{E_c(k-w)E_c(k)}}-\frac{\mathbf{k\cdot w}}{2\sqrt{E_c(k-w)E_c(k)}[E_c(k)+m_c]}\right\}\\
&\times\frac{1}{\mathcal{N}^{J/\psi}}\exp\bigg[\frac{-(\mathbf{k}-\mathbf{w}/2)^2-E^2_c(k)}{\omega^2_{J/\psi}}\bigg]\Psi^{D_l^*}(\mathbf{k}),
\end{split}
\end{equation}
where $\mathcal{N}^{J/\psi}$ is normalization and $\omega_{J/\psi}$=0.826 GeV was obtained from lattice QCD (see details in Ref. \cite{mypaper6}). Then the interaction kernel from one heavy meson ($J/\psi$) exchange has been obtained. The total interaction kernel between $D^{*+}$ and $\bar D^{*0}$ derived from one light meson ($\pi^0$, $\eta$, $\sigma$, $\rho^0$ and $\omega$) exchange and one heavy meson ($J/\psi$) exchange is expressed as Eq. (\ref{kernel1}).

\section{Solution of BS equation in instantaneous approximation}\label{app3}
Here, we give the procedure for solving BS equation expressed as Eq. (\ref{BSE}) with the kernel (\ref{kernel1}) in instantaneous approximation and obtain mass $M_0$ and BS wave function of bound state $D^{*+}\bar D^{*0}$. Substituting BS wave function given by Eq. (\ref{BSWF1+}) and the kernel (\ref{kernel1}) into BS equation (\ref{BSE}), we find that the integral of one term on the right-hand side of Eq. (\ref{BSWF1+}) has contribution to the one of itself and the other terms. Moreover, the cross terms contain the factors of $1/M_1^2$ and $1/M_2^2$, which are small for the masses of heavy mesons are large. It is difficult to strictly solve BS equation, and we will use a simple approach to solve it as follows.

Firstly, we ignore the small cross terms and obtain five individual equations:
\begin{equation}\label{BSE1}
\mathcal{G}^1_{\lambda\tau}(P\cdot p,p^2)=-\int\frac{d^4p'}{(2\pi)^4}\Delta_{F\lambda\theta}(p_1')\mathcal{V}_{\theta\theta',\kappa'\kappa}(p,p';P)\mathcal{G}^1_{\theta'\kappa'}(P\cdot p',p'^2)\Delta_{F\kappa\tau}(p_2'),
\end{equation}
\begin{equation}\label{BSE2}
\mathcal{G}^2_{\lambda\tau}(P\cdot p,p^2)=-\int\frac{d^4p'}{(2\pi)^4}\Delta_{F\lambda\theta}(p_1')\mathcal{V}_{\theta\theta',\kappa'\kappa}(p,p';P)\mathcal{G}^2_{\theta'\kappa'}(P\cdot p',p'^2)\Delta_{F\kappa\tau}(p_2'),
\end{equation}
\begin{equation}\label{BSE3}
\mathcal{G}^3_{\lambda\tau}(P\cdot p,p^2)=-\int\frac{d^4p'}{(2\pi)^4}\Delta_{F\lambda\theta}(p_1')\mathcal{V}_{\theta\theta',\kappa'\kappa}(p,p';P)\mathcal{G}^3_{\theta'\kappa'}(P\cdot p',p'^2)\Delta_{F\kappa\tau}(p_2'),
\end{equation}
\begin{equation}\label{BSE4}
\mathcal{G}^4_{\lambda\tau}(P\cdot p,p^2)=-\int\frac{d^4p'}{(2\pi)^4}\Delta_{F\lambda\theta}(p_1')\mathcal{V}_{\theta\theta',\kappa'\kappa}(p,p';P)\mathcal{G}^4_{\theta'\kappa'}(P\cdot p',p'^2)\Delta_{F\kappa\tau}(p_2'),
\end{equation}
\begin{equation}\label{BSE5}
\mathcal{G}^5_{\lambda\tau}(P\cdot p,p^2)=-\int\frac{d^4p'}{(2\pi)^4}\Delta_{F\lambda\theta}(p_1')\mathcal{V}_{\theta\theta',\kappa'\kappa}(p,p';P)\mathcal{G}^5_{\theta'\kappa'}(P\cdot p',p'^2)\Delta_{F\kappa\tau}(p_2'),
\end{equation}
where $\mathcal{G}^1_{\lambda\tau}(P\cdot p,p^2)=\eta_{\mu}p_{\mu}\epsilon_{\lambda\tau\xi\zeta}p_\xi P_\zeta\mathcal{G}_1(P\cdot p,p^2)$, $\mathcal{G}^2_{\lambda\tau}(P\cdot p,p^2)=\eta_{\mu}\mathcal{T}^7_{\mu\lambda\tau}\mathcal{G}_2(P\cdot p,p^2)$, $\mathcal{G}^3_{\lambda\tau}(P\cdot p,p^2)=\eta_{\mu}\mathcal{T}^8_{\mu\lambda\tau}\mathcal{G}_3(P\cdot p,p^2)$, $\mathcal{G}^4_{\lambda\tau}(P\cdot p,p^2)=\eta_{\mu}\mathcal{T}^9_{\mu\lambda\tau}\mathcal{G}_4(P\cdot p,p^2)$ and $\mathcal{G}^5_{\lambda\tau}(P\cdot p,p^2)=\eta_{\mu}\mathcal{T}^{10}_{\mu\lambda\tau}\mathcal{G}_5(P\cdot p,p^2)$. Comparing the tensor structures in both sides of Eqs. (\ref{BSE1}), (\ref{BSE2}), (\ref{BSE3}), (\ref{BSE4}) and (\ref{BSE5}), respectively, we obtain five individual equations
\begin{equation}\label{BSEGi}
\begin{split}
\mathcal{G}_i(P\cdot p,p^2)=\frac{1}{p_1'^2+M_1^2-i\varepsilon}\frac{1}{p_2'^2+M_2^2-i\varepsilon}\int\frac{d^4p'}{(2\pi)^4}V^{1^+}_{i}(p,p';P)\mathcal{G}_i(P\cdot p',p'^2),
\end{split}
\end{equation}
where $\mathcal{G}_i(P\cdot p,p^2)(i=1,2,3,4,5)$ are scalar functions and $V^{1^+}_{i}(p,p';P)$ are derived from the interaction kernel between $D^{*+}$ and $\bar D^{*0}$.

From Eq. (\ref{ffp}), we can obtain that one light pseudoscalar meson exchange has no contribution to the interaction kernel between two heavy vector mesons for the vector-vector bound state with spin-parity quantum numbers $1^+$. In Ref. \cite{mypaper9}, we found that the contribution from one heavy meson ($J/\psi$) exchange is very small. In instantaneous approximation, we set the momentum of exchanged meson as $w=(\mathbf{w},0)$. Then Eqs. (\ref{BSEGi}) become five relativistic Schr$\ddot{o}$dinger-like equations (see details in Refs. \cite{mypaper4,mypaper5})
\begin{equation}\label{BSEpsii}
\bigg(\frac{b_i^2(M_0)}{2\mu_R}-\frac{\mathbf{p}^2}{2\mu_R}\bigg)\Psi_i^{1^+}(\mathbf{p})=\int\frac{d^3w}{(2\pi)^3}V_{i}^{1^+}(\mathbf{p},\mathbf{w})\Psi_i^{1^+}(\mathbf{p},\mathbf{w}),
\end{equation}
where $\Psi_i^{1^+}(\mathbf{p})=\int dp_0\mathcal{G}_i(P\cdot p,p^2)$, $i=1,2,3,4,5$, $\mu_R=E_1E_2/(E_1+E_2)=[M_0^4-(M_1^2-M_2^2)^2]/(4M_0^3)$, $b^2(M_0)=[M_0^2-(M_1+M_2)^2][M_0^2-(M_1-M_2)^2]/(4M_0^2)$, $E_1=(M_0^2-M_2^2+M_1^2)/(2M_0)$ and $E_2=(M_0^2-M_1^2+M_2^2)/(2M_0)$. The potentials between $D^{*+}$ and $\bar D^{*0}$ up to the second order of the $p/M_{D^*}$ expansion are
\begin{equation}
\begin{split}
V_{1}^{1^+}(\mathbf{p},\mathbf{w})=&-F_1({\mathbf{w}^{2}})\frac{g_\sigma^2}{w^2+M_\sigma^2}F_1({\mathbf{w}^{2}})+F^{(\text{lv})}_2({\mathbf{w}^{2}})F^{(\text{lv})}_2({\mathbf{w}^{2}})\\
&\times\bigg(\frac{-g_\rho^2}{w^2+M_\rho^2}+\frac{g_\omega^2}{w^2+M_\omega^2}\bigg)\bigg(-1-\frac{4\mathbf{p}^2+\mathbf{w}^2}{4E_1E_2}\bigg),
\end{split}
\end{equation}
\begin{equation}
\begin{split}
V_{2}^{1^+}(\mathbf{p},\mathbf{w})=&-F_1({\mathbf{w}^{2}})\frac{g_\sigma^2}{w^2+M_\sigma^2}F_1({\mathbf{w}^{2}})\bigg(1-\frac{\mathbf{w}^2}{M_2^2}\bigg)+F^{(\text{lv})}_2({\mathbf{w}^{2}})F^{(\text{lv})}_2({\mathbf{w}^{2}})\\
&\times\bigg(\frac{-g_\rho^2}{w^2+M_\rho^2}+\frac{g_\omega^2}{w^2+M_\omega^2}\bigg)\bigg(-1+\frac{\mathbf{w}^2}{4E_1E_2}-\frac{2\mathbf{p}^2+\mathbf{w}^2}{2M_2^2}\bigg),
\end{split}
\end{equation}
\begin{equation}
\begin{split}
V_{3}^{1^+}(\mathbf{p},\mathbf{w})&=-F_1({\mathbf{w}^{2}})\frac{g_\sigma^2}{w^2+M_\sigma^2}F_1({\mathbf{w}^{2}})\bigg(1-\frac{\mathbf{w}^2}{M_2^2}\bigg)+F^{(\text{lv})}_2({\mathbf{w}^{2}})F^{(\text{lv})}_2({\mathbf{w}^{2}})\\
&\times\bigg(\frac{-g_\rho^2}{w^2+M_\rho^2}+\frac{g_\omega^2}{w^2+M_\omega^2}\bigg)\bigg(-1-\frac{4\mathbf{p}^2+5\mathbf{w}^2}{4E_1E_2}+\frac{\mathbf{w}^2}{M_2^2}\bigg),
\end{split}
\end{equation}
\begin{equation}
\begin{split}
V_{4}^{1^+}(\mathbf{p},\mathbf{w})&=-F_1({\mathbf{w}^{2}})\frac{g_\sigma^2}{w^2+M_\sigma^2}F_1({\mathbf{w}^{2}})+F^{(\text{lv})}_2({\mathbf{w}^{2}})F^{(\text{lv})}_2({\mathbf{w}^{2}})\\
&\times\bigg(\frac{-g_\rho^2}{w^2+M_\rho^2}+\frac{g_\omega^2}{w^2+M_\omega^2}\bigg)\bigg(-1-\frac{4\mathbf{p}^2+\mathbf{w}^2}{4E_1E_2}+\frac{\mathbf{p}^2}{M_2^2}\bigg),
\end{split}
\end{equation}
\begin{equation}
\begin{split}
V_{5}^{1^+}(\mathbf{p},\mathbf{w})&=-F_1({\mathbf{w}^{2}})\frac{g_\sigma^2}{w^2+M_\sigma^2}F_1({\mathbf{w}^{2}})+F^{(\text{lv})}_2({\mathbf{w}^{2}})F^{(\text{lv})}_2({\mathbf{w}^{2}})\\
&\times\bigg(\frac{-g_\rho^2}{w^2+M_\rho^2}+\frac{g_\omega^2}{w^2+M_\omega^2}\bigg)\bigg(-1-\frac{4\mathbf{p}^2+\mathbf{w}^2}{4E_1E_2}\bigg).
\end{split}
\end{equation}
Solving these five equations (\ref{BSEpsii}), respectively, one can obtain the eigenvalues $b_i^2(M_0)$ and the corresponding eigenfunctions $\Psi_i^{1^+}(\mathbf{p})$. From $\Psi_i^{1^+}$, it is easy to obtain $\mathcal{G}_i$, respectively.

Next, we consider the cross terms in BS equation. Because the cross terms are small, we can take the ground state BS wave function to be a linear combination of five eigenstates $\mathcal{G}^{i0}_{\lambda\tau}(P\cdot p,p^2)(i=1,2,3,4,5)$ corresponding to lowest energy in Eqs. (\ref{BSE1}), (\ref{BSE2}), (\ref{BSE3}), (\ref{BSE4}) and (\ref{BSE5}). Then in the basis provided by $\mathcal{G}^{10}_{\lambda\tau}(P\cdot p,p^2)=\eta_{\mu}p_{\mu}\epsilon_{\lambda\tau\xi\zeta}p_\xi P_\zeta\mathcal{G}_{10}(P\cdot p,p^2)$, $\mathcal{G}^{20}_{\lambda\tau}(P\cdot p,p^2)=\eta_{\mu}\mathcal{T}^7_{\mu\lambda\tau}\mathcal{G}_{20}(P\cdot p,p^2)$, $\mathcal{G}^{30}_{\lambda\tau}(P\cdot p,p^2)=\eta_{\mu}\mathcal{T}^8_{\mu\lambda\tau}\mathcal{G}_{30}(P\cdot p,p^2)$, $\mathcal{G}^{40}_{\lambda\tau}(P\cdot p,p^2)=\eta_{\mu}\mathcal{T}^9_{\mu\lambda\tau}\mathcal{G}_{40}(P\cdot p,p^2)$ and $\mathcal{G}^{50}_{\lambda\tau}(P\cdot p,p^2)=\eta_{\mu}\mathcal{T}^{10}_{\mu\lambda\tau}\mathcal{G}_{50}(P\cdot p,p^2)$, BS wave function $\chi^{1^+}_{\lambda\tau}$ is considered as
\begin{equation}
\begin{split}\label{BSwfapprox}
\chi_{\lambda\tau}^{1^+}(P,p)=&\frac{1}{\mathcal{N}^{1^+}}\eta^{\varsigma=1,2,3}_{\mu}(P)\chi_{\mu\lambda\tau}(P,p)\\
=&\frac{1}{\mathcal{N}^{1^+}}\eta^\varsigma_{\mu}(P)(\mathcal{C}_1p_{\mu}\epsilon_{\lambda\tau\xi\zeta}p_\xi P_\zeta\mathcal{G}_{10}+\mathcal{C}_2\mathcal{T}^7_{\mu\lambda\tau}\mathcal{G}_{20}+\mathcal{C}_3\mathcal{T}^8_{\mu\lambda\tau}\mathcal{G}_{30}+\mathcal{C}_4\mathcal{T}^{9}_{\mu\lambda\tau}\mathcal{G}_{40}+\mathcal{C}_5\mathcal{T}^{10}_{\mu\lambda\tau}\mathcal{G}_{50}).
\end{split}
\end{equation}
Substituting (\ref{BSwfapprox}) into BS equation (\ref{BSE}) and comparing the tensor structures in both sides, we obtain an eigenvalue equation in instantaneous approximation \cite{mypaper4}
\begin{equation}\label{eigeneq}
\left(\begin{array}{ccccc}
\frac{b_{10}^2(M_0)}{2\mu_R}-\lambda&0&0&0&0\\
0&\frac{b_{20}^2(M_0)}{2\mu_R}-\lambda&0&0&H_{25}\\
0&0&\frac{b_{30}^2(M_0)}{2\mu_R}-\lambda&H_{34}&0\\
0&0&H_{43}&\frac{b_{40}^2(M_0)}{2\mu_R}-\lambda&0\\
0&H_{52}&0&0&\frac{b_{50}^2(M_0)}{2\mu_R}-\lambda\end{array}\right)\left(\begin{array}{c}\mathcal{C}'_1\\\mathcal{C}'_2\\\mathcal{C}'_3\\\mathcal{C}'_4\\\mathcal{C}'_5\end{array}\right)=0,
\end{equation}
where we have the matrix elements
\begin{equation}
\begin{split}
H_{25}=H_{52}=&\int \frac{d^3p}{(2\pi)^3}\Psi_{20}^{1^+}(\mathbf{p})^*\int\frac{d^3w}{(2\pi)^3}\bigg[F_1({\mathbf{w}^{2}})\frac{g_\sigma^2}{w^2+M_\sigma^2}F_1({\mathbf{w}^{2}})\frac{\mathbf{w}^2}{2M_2^2}+F^{(\text{lv})}_2({\mathbf{w}^{2}})F^{(\text{lv})}_2({\mathbf{w}^{2}})\\
&\times\bigg(\frac{-g_\rho^2}{w^2+M_\rho^2}+\frac{g_\omega^2}{w^2+M_\omega^2}\bigg)\bigg(\frac{2\mathbf{p}^2+3\mathbf{w}^2}{4E_1E_2}-\frac{2\mathbf{p}^2+\mathbf{w}^2}{4M_2^2}\bigg)\bigg]\Psi_{50}^{1^+}(\mathbf{p},\mathbf{w}),
\end{split}
\end{equation}
\begin{equation}
\begin{split}
H_{34}=H_{43}=&\int \frac{d^3p}{(2\pi)^3}\Psi_{30}^{1^+}(\mathbf{p})^*\int\frac{d^3w}{(2\pi)^3}\bigg[F_1({\mathbf{w}^{2}})\frac{g_\sigma^2}{w^2+M_\sigma^2}F_1({\mathbf{w}^{2}})\frac{\mathbf{w}^2}{2M_2^2}\\
&+F^{(\text{lv})}_2({\mathbf{w}^{2}})F^{(\text{lv})}_2({\mathbf{w}^{2}})\bigg(\frac{-g_\rho^2}{w^2+M_\rho^2}+\frac{g_\omega^2}{w^2+M_\omega^2}\bigg)\frac{\mathbf{w}^2}{2M_2^2}\bigg]\Psi_{40}^{1^+}(\mathbf{p},\mathbf{w}),
\end{split}
\end{equation}
and $b_{i0}^2(M_0)/(2\mu_R)(i=1,2,3,4,5)$ are the eigenvalues corresponding to lowest energy in Eqs. (\ref{BSEpsii}), respectively; $\Psi_{i0}^{1^+}$ are the corresponding eigenfunctions. From this equation, we can obtain the eigenvalues and eigenfunctions which contain the contribution from the cross terms. Equations (\ref{BSEpsii}) can be solved numerically with these form factors, and then the eigenvalue equation (\ref{eigeneq}) can be solved. The mass $M_0$ and BS wave function of bound state $D^{*+}\bar{D}^{*0}$ with spin-parity quantum numbers $1^+$ can be obtained.

\section{BS wave function for the heavy axial-vector meson $h_c(1P)$}\label{app4}
According to the most general form of Bethe-Salpeter wave functions for mesons of arbitrary spin and definite parity expressed as Eq. (13) in Ref. \cite{BSWF:Dai}, Bethe-Salpeter wave function for heavy axial-vector meson $h_c(1P)$ has the form
\begin{equation}
\begin{split}\label{hcBSWF1}
\chi_{5\nu}(Q,q)=&\frac{1}{\mathcal{N}^{h_c}}\bigg\{\gamma_5\bigg(q_\nu+Q_\nu\frac{Q\cdot q}{M_{h_c}^2}\bigg)\varphi_{h_c1}(Q\cdot q,q^2)+i\gamma_5\bigg(q_\nu+Q_\nu\frac{Q\cdot q}{M_{h_c}^2}\bigg)\gamma\cdot Q\varphi_{h_c2}(Q\cdot q,q^2)\\
&+\gamma_5\bigg[\bigg(\gamma_\nu+Q_\nu\frac{\gamma\cdot Q}{M_{h_c}^2}\bigg)\gamma_\upsilon q_\upsilon-\gamma_\upsilon q_\upsilon\bigg(\gamma_\nu+Q_\nu\frac{\gamma\cdot Q}{M_{h_c}^2}\bigg)\bigg]\varphi_{h_c3}(Q\cdot q,q^2)\\
&+i\epsilon_{\nu\nu'\upsilon\upsilon'}\gamma_{\nu'}q_\upsilon Q_{\upsilon'}\varphi_{h_c4}(Q\cdot q,q^2)\bigg\},
\end{split}
\end{equation}
where $Q$ is the momentum of $h_c(1P)$ meson, $q$ becomes the relative momentum between $c$-quark and $c$-antiquark, $\chi_{5\nu}(Q,q)$ is transverse ($Q_{\nu}\chi_{5\nu}(Q,q)=0$), and $\varphi_{h_ci}(Q\cdot q,q^2)(i=1,2,3,4)$ are independent scalar functions. In fact, $4$-vectors $q_\nu$ and $q_{\upsilon}$ represent the orbital angular momenta, and for $P$-wave state there is a $4$-vector $q_\nu$ or $q_{\upsilon}$ in each term of BS wave function (\ref{hcBSWF1}). The heavy axial-vector meson $h_c(1P)$ with equal-mass constituents has quantum numbers $1^{+-}$. In Eq. (\ref{hcBSWF1}), scalar functions $\varphi_{h_c1}(Q\cdot q,q^2)$ and $\varphi_{h_c2}(Q\cdot q,q^2)$ should be even functions of $Q\cdot q$, and $\varphi_{h_c3}(Q\cdot q,q^2)$ and $\varphi_{h_c4}(Q\cdot q,q^2)$ should be odd functions of $Q\cdot q$ \cite{BSE:Krassnigg}. Since $\varphi_{h_ci}(Q\cdot q,q^2)$ depend on $q^2$ and $Q\cdot q$, we assume that in Euclidean space these scalar functions have the explicit form
\begin{equation}
\begin{split}
\varphi_{h_c1}(Q\cdot q,q^2)&=\frac{1}{(Q\cdot q)^2-\mathcal{C}_{\omega_1}}\exp\bigg(-\frac{q^2}{\omega^2_{h_c1}}\bigg),\\
\varphi_{h_c2}(Q\cdot q,q^2)&=\frac{1}{M_{h_c}}\frac{1}{(Q\cdot q)^2-\mathcal{C}_{\omega_2}}\exp\bigg(-\frac{q^2}{\omega^2_{h_c2}}\bigg),\\
\varphi_{h_c3}(Q\cdot q,q^2)&=M_{h_c}^2\frac{Q\cdot q}{[(Q\cdot q)^2-\mathcal{C}_{\omega_3}]^2}\exp\bigg(-\frac{q^2}{\omega^2_{h_c3}}\bigg),\\
\varphi_{h_c4}(Q\cdot q,q^2)&=M_{h_c}\frac{Q\cdot q}{[(Q\cdot q)^2-\mathcal{C}_{\omega_4}]^2}\exp\bigg(-\frac{q^2}{\omega^2_{h_c4}}\bigg).
\end{split}
\end{equation}

The effective theory at low energy QCD only involves the light quarks ($u,d,s$) and the effective interaction Lagrangian can not involve the heavy quark $c$, expressed as Eq. (\ref{Lag}). For heavy meson, we have to consider that heavy meson is a bound state composed of a quark and an antiquark and use perturbative QCD to deal with $c\bar c$ bound state. We employ the gluon propagator which includes the information of condensates
\begin{equation}
\begin{split}\label{gluonpro}
D^{a'b'}_{\rho\sigma}(s)&=\delta^{a'b'}D_{\rho\sigma}(s)=\delta^{a'b'}\delta_{\rho\sigma}\frac{-i}{s^2-4\pi\alpha_ss^2C(s^2)},
\end{split}
\end{equation}
where $s$ represents the gluon momentum, superscripts $a'$ and $b'$ are color indices, and $\alpha_s$ is the quark-gluon coupling constant. In QCD sum rules, the analytic form for function $C(s^2)$ has been obtained \cite{cc2}
\begin{equation}
\begin{split}
C(s^2)&=\bigg\langle\frac{\alpha_s}{\pi}G^2\bigg\rangle\frac{1}{48}\frac{1}{s^4}\bigg[\frac{3(1+u^2)(1-u^2)^2}{2u^5}\text{ln}\frac{1+u}{1-u}-\frac{3u^4-2u^2+3}{u^4}\bigg],\\
u&=(1+4m_cm_c/s^2)^{1/2},
\end{split}
\end{equation}
where $\langle\frac{\alpha_s}{\pi}G^2\rangle$ is the gluon condensate. Then we can obtain BS equation for heavy axial-vector meson $h_c(1P)$ in ladder approximation
\begin{equation}
\begin{split}\label{BSEhc}
\chi_{5\nu}(Q,q)=\int\frac{d^4q'}{(2\pi)^4}\frac{4}{3}4\pi\alpha_sS_F(q+Q/2)\gamma_\rho D_{\rho\sigma}(s)\chi_{5\nu}(Q,q')\gamma_\sigma S_F(q-Q/2),
\end{split}
\end{equation}
where $s=q'-q$, $S_F(p)=-1/(\gamma\cdot p-im_c)$ and color indices have been calculated.

In this work, we employ the coupling constant $\alpha_s=0.20$, the gluon condensate $\langle\frac{\alpha_s}{\pi}G^2\rangle=(0.36~\text{GeV})^4$ \cite{cc2}, the heavy quark mass $m_{c}=1.55$ GeV \cite{ts:Ebert} and the heavy axial-vector meson mass $M_{h_c}=3.525$ GeV \cite{PDG2024}. Substituting BS wave function given by Eq. (\ref{hcBSWF1}) and the gluon propagator (\ref{gluonpro}) into BS equation (\ref{BSEhc}) and fitting scalar functions in both sides of this BS equation, we find that the first and second terms on the right-hand side of Eq. (\ref{hcBSWF1}) are dominant and other terms are almost irrelevant. Then Bethe-Salpeter wave function for heavy axial-vector meson $h_c(1P)$ can be considered as
\begin{equation}
\begin{split}\label{hcBSWF2}
\chi_{5\nu}(Q,q)&=\frac{1}{\mathcal{N}^{h_c}}\bigg[\gamma_5\bigg(q_\nu+Q_\nu\frac{Q\cdot q}{M_{h_c}^2}\bigg)\varphi_{h_c1}(Q\cdot q,q^2)+i\gamma_5\bigg(q_\nu+Q_\nu\frac{Q\cdot q}{M_{h_c}^2}\bigg)\gamma\cdot Q\varphi_{h_c2}(Q\cdot q,q^2)\bigg].
\end{split}
\end{equation}
By fitting scalar functions in both sides of equation (\ref{BSEhc}), we obtain $\mathcal{C}_{\omega_1}=(0.86~\text{GeV})^4$, $\omega_{h_c1}$=0.89 GeV, $\mathcal{C}_{\omega_2}=(0.88~\text{GeV})^4$ and $\omega_{h_c2}$=0.65 GeV.

\bibliographystyle{apsrev}
\bibliography{ref}

\end{document}